\documentclass[fleqn,usenatbib]{mnras}

\usepackage{savesym}
\usepackage{graphicx}
\usepackage{newtxtext,newtxmath}
\usepackage{bm}
\savesymbol{Bbbk}
\usepackage{amssymb}  
\savesymbol{Bbbk}
\usepackage{dsfont}  
\usepackage{xcolor}  
\usepackage[capitalise, nameinlink, noabbrev]{cleveref} 
\usepackage{tikz} 
\usetikzlibrary{shapes.geometric, arrows}

\tikzstyle{startstop} = [rectangle, rounded corners, minimum width=3cm, minimum height=1cm,text centered, draw=black, fill=red!30]
\tikzstyle{io} = [trapezium, trapezium left angle=70, trapezium right angle=110, minimum width=3cm, minimum height=1cm, text centered, draw=black, fill=blue!30]
\tikzstyle{process} = [rectangle, minimum width=3cm, minimum height=1cm, text centered, draw=black, fill=orange!30]
\tikzstyle{decision} = [diamond, minimum width=3cm, minimum height=1cm, text centered, draw=black, fill=green!30]
\tikzstyle{arrow} = [thick,->,>=stealth]

\creflabelformat{equation}{#2#1#3}

\DeclareMathOperator{\erf}{erf}

\title[Spatially Offset BHs in Horizon-AGN]{Spatially offset black holes in the Horizon-AGN simulation and comparison to observations}

\author[D. J. Bartlett et al.]{Deaglan J. Bartlett,$^{1}$\thanks{E-mail: deaglan.bartlett@physics.ox.ac.uk}
Harry Desmond$^{1}$,
Julien Devriendt$^{1}$,
Pedro G. Ferreira$^{1}$\newauthor 
and Adrianne Slyz$^{1}$
\\
$^{1}$Astrophysics, University of Oxford, Denys Wilkinson Building, Keble Road, Oxford, OX1 3RH, UK
}

\date{Accepted XXX. Received YYY; in original form ZZZ}
\pubyear{2020}

\begin{document}
\label{firstpage}
\pagerange{\pageref{firstpage}--\pageref{lastpage}}
\maketitle

\begin{abstract}
    We study the displacements between the centres of galaxies and their supermassive black holes (BHs) in the cosmological hydrodynamical simulation Horizon-AGN, and in a variety of observations from the literature. The BHs in Horizon-AGN feel a sub-grid dynamical friction force, sourced by the surrounding gas, which prevents recoiling BHs being ejected from the galaxy. We find that i) the fraction of spatially offset BHs increases with cosmic time, ii) BHs live on prograde orbits in the plane of the galaxy with an orbital radius that decays with time but stalls near $z=0$, and iii) the magnitudes of offsets from the galaxy centres are substantially larger in the simulation than in observations. We attribute the stalling of the infall and excessive offset magnitudes to the fact that dynamical friction from stars and dark matter is not modelled in the simulation, and hence provide a way to improve the black hole dynamics of future simulations.
\end{abstract}


\section{Introduction}

Most galaxies are now known to harbour supermassive black holes (SMBHs) near their centres. The strong correlations between SMBH mass and galactic properties such as velocity dispersion \citep{Kormendy_2000, Gebhardt_2000,Ferrarese_2000} and stellar and bulge mass \citep{Magorrian_1998,Marleau_2013} show that they are far from passive onlookers in their hosts' evolution, but rather play an key role in shaping the galaxy population. 

The high bolometric luminosities of active galactic nuclei (AGNs) are due to accretion of matter onto SMBHs. Tens of thousands of AGNs have been discovered in the central regions of galaxies, although the precise location of the BH need not be coincident with the galactic centre. Upon the merger of two galaxies a BH binary may form near the centre of the merged system, which can coalesce due to stellar and gaseous interactions \citep{Begelman_1980}. Gravitational wave emission upon coalescence can, by linear momentum conservation, cause the centre of mass to recoil \citep{Peres_1962,Bekenstein_1973}, thus resulting in a BH offset from the galactic centre. It should be possible to observe the coalescence of these binaries in the early Universe with the \textit{Laser Interferometer Space Antenna (LISA)} \citep{Sesana_2004}. Other processes can result in this phenomenon such as three-body interactions between BHs if there are two successive mergers \citep{Hut_1992,Xu_1994}, or subhalo accretion, which transfers energy to the BH by dynamical friction, resulting in offsets of tens of parsecs \citep{boldrini2020subhalo}. A population of offset and wandering BHs \citep{Volonteri_2003} is therefore expected.

The same three-body interactions that can physically eject BHs from galaxies can result in numerical artefacts in cosmological simulations through numerical heating of BH particles \citep{Hernquist_1990}. This is especially prevalent when the BH and dark matter particles have similar masses. Many simulations implement expedient but unphysical schemes to alleviate this problem, such as `teleporting' the BH back to the local potential minimum at each time step. This is not the case however in the Horizon-AGN simulation, which uses a more physically motivated model -- dynamical friction between the BH and surrounding gas -- to impose a drag force on recoiling BHs \citep{Dubois_2014,Volonteri_2016}. By studying the properties of offset BHs in this simulation, we develop both a physical picture of the evolution of individual systems and a statistical sense of the behaviour of the entire population. This will enable us to assess the efficacy of the dynamical friction model.

Knowledge of the location of central BHs is essential for models of galaxy formation.
The number of wandering BHs depends on the degree of dissipation in galaxy mergers, which in turn determines the BHs' evolution through, and hence scatter around, the $M_{\rm BH}-\sigma$ relation \citep{Kazantzidis_2004}. The lower gas densities around an off-centre BH restricts BH accretion, which could result in a lower mass BH than one constrained to reside at the centre \citep{Tremmel_2015}. The lower accretion rate would quench BH feedback, and hence reduce the impact on the surrounding gas, stars and dark matter. This is particularly important in attempts to resolve small-scale problems in $\Lambda$CDM such as the core-cusp problem through BH feedback \citep{Pontzen_2012, boldrini2020subhalo}. The location of SMBHs is also important to the study of dark matter microphysics: self-interacting dark matter (SIDM) \citep{Spergel_2000,Burket_2000}, for example, lowers central dark matter densities in halos and hence lengthens dynamical friction timescales and increases the fraction of off-centre BHs \citep{diCintio_2017}. SIDM also suppresses BH growth and feedback, allowing higher star formation rates in $\Lambda$SIDM galaxies than their $\Lambda$CDM counterparts \citep{Cruz_2020}.

The structure of this paper is as follows. In \autoref{sec:Horizon-AGN} we introduce the Horizon-AGN simulation, and in \autoref{sec:Observational Data} we summarise the observational datasets against which we will compare the simulation results. We outline the methods used to make this comparison in \autoref{sec:Comparing Horizon-AGN with data} and present the results in \autoref{sec:Results}. These are discussed in \autoref{sec:Discussion} and our conclusions are given in \autoref{sec:Conclusions}.

\section{Horizon-AGN}
\label{sec:Horizon-AGN}

Horizon-AGN\footnote{\url{http://www.horizon-simulation.org/about.html}} \citep{Dubois_2014} is a large-volume cosmological hydrodynamical simulation, run with the Adaptive Mesh Refinement code, \textsc{ramses} \citep{Teyssier_2002}. The adopted standard $\Lambda$CDM cosmology is compatible with a WMAP-7 cosmology \citep{Komatsu_2011} and thus has total matter density $\Omega_{\rm m} = 0.272$, dark energy density $\Omega_{\rm \Lambda} = 0.728$, amplitude of the matter power spectrum $\sigma_8 = 0.81$, baryon density $\Omega_{\rm b} = 0.045$, Hubble constant $H_0 = 70.4 {\rm \, km \, s^{-1} \, Mpc^{-1}}$, and power-spectrum slope $n_{\rm s} = 0.967$. Throughout this paper we use WMAP-7 parameters, to remain consistent with Horizon-AGN. The size of the box is $L_{\rm box} = 100 h^{-1} {\rm \, Mpc}$ and contains $1024^3$ DM particles, resulting in a dark matter (DM) mass resolution of $M_{\rm DM} = 8 \times 10^7 {\rm M_{\sun}}$.

The simulation incorporates prescriptions for background UV heating, gas cooling (including the contribution from metals released by stellar feedback) and feedback from stellar winds and type Ia and type II supernovae assuming a Salpeter initial mass function (IMF) \citep{Dubois_2008,taysun2012a}. Star formation follows a Schmidt law with a 1 per cent efficiency \citep{Rasera_2006} and a star formation density threshold of $n_0 = 0.1 {\rm \, H \, cm^{-3}}$.

A cell is refined up to an effective physical resolution of $\Delta x = 1 {\rm \, kpc}$, with a new refinement level added if the mass in a cell is more than 8 times that of the initial mass resolution. The force softening scale is $\sim 2 {\rm \, kpc}$.

\subsection{Black hole formation, growth and feedback on ambient gas}

If the combined gas and stellar density exceeds the threshold for star formation in a cell and if the stellar velocity dispersion within that cell surpasses $100 {\rm \, km \, s^{-1}}$, a black hole (BH) is created with an initial seed mass of $10^5 {\rm \, M_{\sun}}$. A BH cannot form if this occurs within 50 comoving ${\rm kpc}$ of another BH, preventing many BHs from forming in the same galaxy. BHs grow through mergers and accretion, with the accretion rate given by the Bondi-Hoyle-Littleton rate multiplied by a dimensionless boost factor \citep{Booth_2009}
\begin{equation}
	\alpha = 
		\begin{cases}
			\mbox{$\left(\rho_{\rm gas} / \rho_0 \right)^2$} & \text{if} \quad \rho_{\rm gas} > \rho_0 \\
			\mbox{$1$} & \text{otherwise},
		\end{cases}
\end{equation}
for gas density $\rho_{\rm gas}$, and is capped at the Eddington rate with an assumed radiative efficiency of $\epsilon_{\rm r} = 0.1$ for the \citet{Shakura_1973} accretion onto a Schwarzschild BH. $\rho_0$ is the mass density of hydrogen if its number density is $n_0 = 0.1 {\rm H \, cm^{-3}}$. This boost factor accounts for the inability to model the colder and higher density regions of the interstellar medium.

The AGN feedback is a combination of two different modes: the `radio mode' for $\chi < 0.01$ and the `quasar mode' otherwise, where
\begin{equation}
	\chi = \frac{\dot{M}_{\rm BH}}{\dot{M}_{\rm Edd}},
\end{equation}
for BH and Eddington accretion rates $\dot{M}_{\rm BH}$ and $\dot{M}_{\rm Edd}$ respectively. The quasar mode isotropically ejects thermal energy with deposition rate $\dot{E}_{\rm AGN} = \epsilon_{\rm f} \epsilon_{\rm r} \dot{M}_{\rm BH} c^2$ into the gas within a sphere of radius $\Delta x$. The efficiency $\epsilon_{\rm f}$ is taken to be 0.15 as this reproduces the correlations between BHs and galaxies and the BH density in our local Universe \citep[see][]{Dubois_2012}. In contrast, the radio mode releases the feedback energy into a bipolar, cylindrical outflow with height $2 \Delta x$ and radius $\Delta x$ as in \citet{Omma_2004}. The jet velocity is $10^4 {\rm \, km \, s^{-1}}$ and the radio mode has an increased efficiency, with $\epsilon_{\rm f} = 1$.3. Note that `radio mode' does not necessarily mean `radio loud'. For a detailed analysis of how the Horizon-AGN simulation compares to observational radio data see \citet{Slyz_2015}.

Due to the finite resolution of the simulation, for the rest of the paper we ignore BHs with masses $M_{\rm B} < 2 \times 10^7 {\rm \, M_{\sun}}$ \citep[see][]{Volonteri_2016}.

\subsection{Dynamical friction}
\label{sec:Dynamical friction}

Simulating the dynamical friction, which  ensures the BH's trajectory decays towards the centre of the galaxy \citep{Chandrasekhar_1943,Binney_2008}, is notoriously challenging for cosmological simulations because the gas cannot be tracked all the way down to the BH \citep{Beckmann_2018}. Some simulations \citep{Taylor_2014,Schaye_2015,Sijacki_2015} therefore anchor the BH to the centre of their DM halos. Offset BHs might not efficiently accrete since high density gas tends to be centrally located \citep{Smith_2018}, so the feedback of such BHs is quenched \citep{boldrini2020subhalo}. Consequently, pinning the BH to the halo centre can result in unrealistic BH and galactic evolution.

Artificial advection schemes exist to overcome this, but these are not without their drawbacks \citep[see][]{Tremmel_2015}. It is necessary to introduce a sub-grid model for the dynamical friction induced by the gas surrounding the BH. In Horizon-AGN dynamical friction is modelled as
\begin{equation}
\label{eq:Dynamical friction}
F_{\rm DF} = f_{\rm gas} 4 \pi \alpha \rho_{\rm gas} \left( \frac{G M_{\rm BH}}{\bar{c}_s} \right)^2,
\end{equation}
where $\bar{c}_s$ is the average sound speed and the coefficient $f_{\rm gas} \in [0, 2]$ is a function of the Mach number $\mathcal{M} = \bar{u} / \bar{c}_s < 1$ \citep{Ostriker_1999,Chapon_2013}, where $\bar{u}$ is the average velocity of the gas relative to the BH. The average density and sound speed around the BH are computed using kernel weighting of neighbouring cell values, whereas the average relative
velocity is set to a constant value typical of gas velocity dispersion in the ISM, $\bar{u}=u_{\rm max}= 10 {\rm \, km \, s^{-1}}$, as described in \citet{Dubois_2012}.

\subsection{Assigning BHs to galaxies}

\begin{figure}
\begin{tikzpicture}[node distance=1.5cm]

\node (pro1) [process] {Get most massive halo not yet considered.};
\node (pro2) [process, below of =pro1, text width = 7cm] {Find all available galaxies with $0.1 r_{\rm vir}$ of halo centre.};
\node (pro3) [process, below of=pro2, text width = 7cm] {Assign most massive of these galaxies to halo and make this galaxy unavailable to other halos.};
\node (pro4) [process, below of=pro3, text width = 7cm] {Find all available BHs within $2r_{\rm eff}$ of galaxy centre and $0.1 r_{\rm vir}$ of halo centre.};
\node (pro5) [process, below of=pro4, text width = 7cm] {Make desired BH selection cuts (e.g. mass, luminosity or accretion efficiency).};
\node (pro6) [process, below of=pro5, text width = 7cm] {Assign central BH to galaxy and halo using desired method (e.g. closest or most massive BH) and make this BH unavailable to other halos and galaxies.};
\node (dec1) [decision, below of=pro6, yshift=-1.cm, text width=2cm] {More haloes to consider?};
\node (stop) [startstop, below of=dec1, yshift=-1.cm] {Stop};

\node [coordinate] (node1) [right of=pro1, xshift=2.5cm] {};
\node [coordinate] (node2) [right of=dec1, xshift=2.5cm] {};

\draw [arrow] (pro1) -- (pro2);
\draw [arrow] (pro2) -- (pro3);
\draw [arrow] (pro3) -- (pro4);
\draw [arrow] (pro4) -- (pro5);
\draw [arrow] (pro5) -- (pro6);
\draw [arrow] (pro6) -- (dec1);
\draw [arrow] (dec1) -- node[anchor=east] {no} (stop);
\draw [thick] (dec1) -- node[anchor=south] {yes} (node2) -- (node1);
\draw [arrow] (node1) -- (pro1);

\end{tikzpicture}
\caption{\label{fig:flowchart}Process of defining a halo--galaxy--BH system in Horizon-AGN from the individual objects. The selection cuts made on the BHs vary depending on the observational sample we compare to (see \autoref{sec:Observational Data}).}
\end{figure}
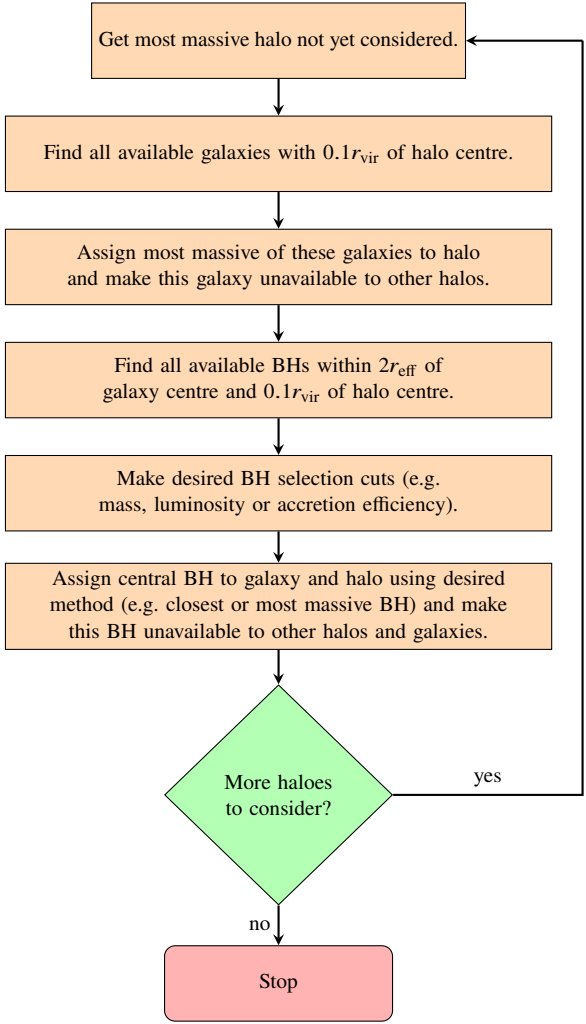

Since the BHs are not labelled by their host halo number in the simulation, we must do this in post-processing. In this section we outline how this is achieved and summarise the procedure in \autoref{fig:flowchart}. Haloes and galaxies are identified using the \textsc{adaptaHOP} structure finder \citep{Aubert_2004,Tweed_2009} applied to dark matter and star particles, respectively. In both cases, we need a minimum of 50 particles and the smoothed density field, calculated using the 20 nearest neighbours, must exceed 178 times the mean total matter density \citep{Gunn_1972}. The centre is taken as the position of the densest particle, after a shrinking sphere approach is used \citep{Power_2003}. Therefore the halo centre is defined to be the position of the densest dark matter particle, and the galactic centre is at the location of the densest star particle.

We look for galaxies within 10 per cent of the virial radius, $r_{\rm vir}$, of a DM halo and match the most massive unassigned galaxy to that halo \citep{Chisari_2017} to produce a galaxy+halo structure. We consider each halo in turn, moving from the most to least massive. A BH is assigned to a galaxy+halo structure if it is within twice the effective radius, $r_{\rm eff}$, of the galaxy and 10 per cent of $r_{\rm vir}$ of the halo. At this point, some of the galaxy+halo structures contain multiple BHs, since the galaxies contain up to 59 BHs within $2 r_{\rm eff}$ of their centres. To decide which is the central BH, we can make one of several choices
\begin{enumerate}
	\item Select the most massive BH \citep[as in][]{Volonteri_2016}.
	\item Ignore any BH with $L < L_{\rm cut}$, where $L = \epsilon_{\rm r}  \dot{M}_{\rm BH} c^2$ and $L_{\rm cut}$ is some cut-off luminosity to be decided. Select the BH from the remaining candidates that is closest to the centre of the galaxy. This is inspired by \citet{Volonteri_2016}, who find that higher luminosity BHs tend to reside closer to galaxies' centres, so we expect this cut to preferentially choose BHs near the centre.
	\item Simply select the BH closest to the centre of the galaxy.
\end{enumerate}

In each case we work hierarchically through the haloes, going from the most to least massive. Once a BH has been assigned to a galaxy+halo structure, it is removed from the list of available BHs.

We can now calculate the distance between the centre of the galaxy and its central BH, $r_{\rm GB}$, using the three dimensional information. We can also project the offsets onto the plane of the sky according to an observer at the centre of the simulation volume, to determine the two-dimensional offset that would be observed.

We start by investigating the effects of the various selection procedures since conclusions made about the offset population are sensitive to the way we assign BHs to galaxies. For the remainder of this section, we use the $z=0.1$ output for Horizon-AGN and consider the three-dimensional offsets.

\begin{figure}
	\centering
	\includegraphics[width=\columnwidth]{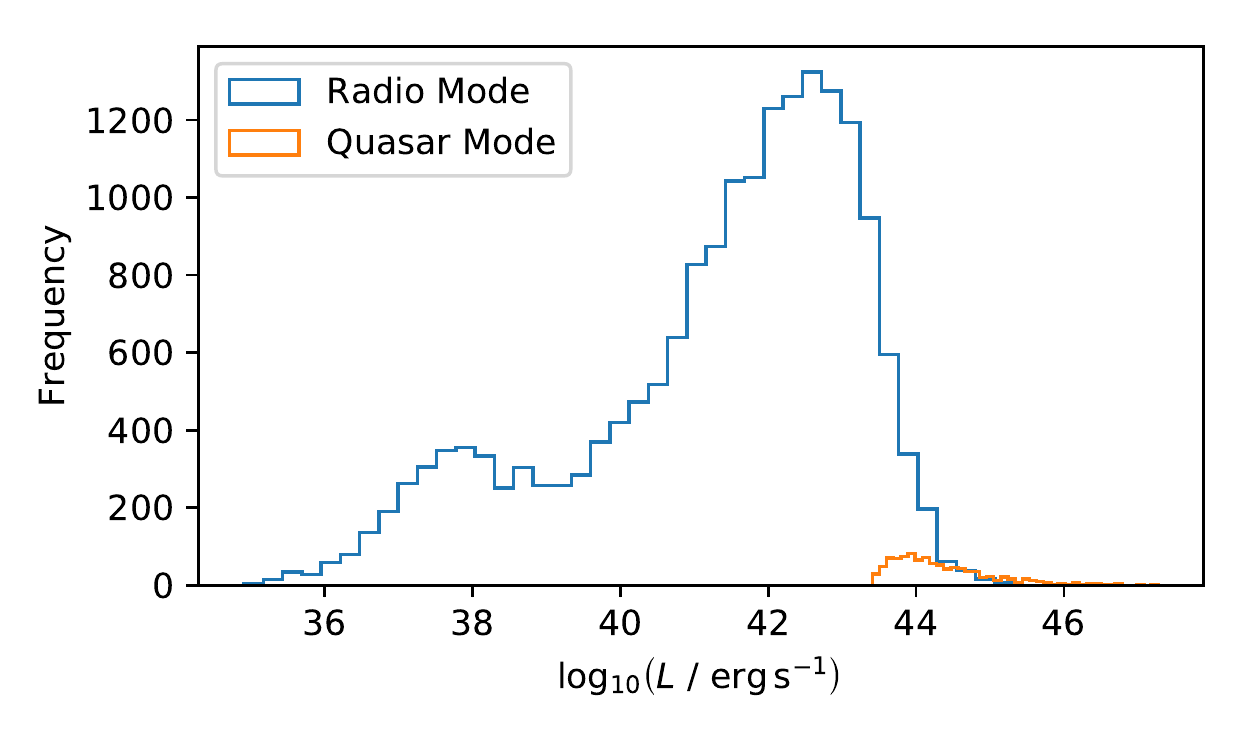}
	\caption{\label{fig:bh_luminosity_distribution} Distribution of the luminosity, $L$, of BHs in Horizon-AGN at $z = 0.1$ with masses $M > 2 \times 10^7 {\rm \, M_{\sun}}$. The distribution is bimodal with peaks at $L \sim 10^{38}$ and $10^{43} {\rm \, erg \, s^{-1}}$. BHs only operate in the quasar mode at very high luminosity and are greatly outnumbered by those in the radio mode.}
\end{figure}

To decide what to use as $L_{\rm cut}$, in \autoref{fig:bh_luminosity_distribution} we plot the BH luminosity distribution. Since we are interested in the high-luminosity region, as we expect high-luminosity BHs to reside near the centre \citep{Volonteri_2016}, we perform two cuts near the peaks of the bimodal distribution, at $L_{\rm cut} = 10^{38}$ and $10^{43} {\rm \, erg \, s^{-1}}$.

\begin{figure}
	\centering
	\includegraphics[width=\columnwidth]{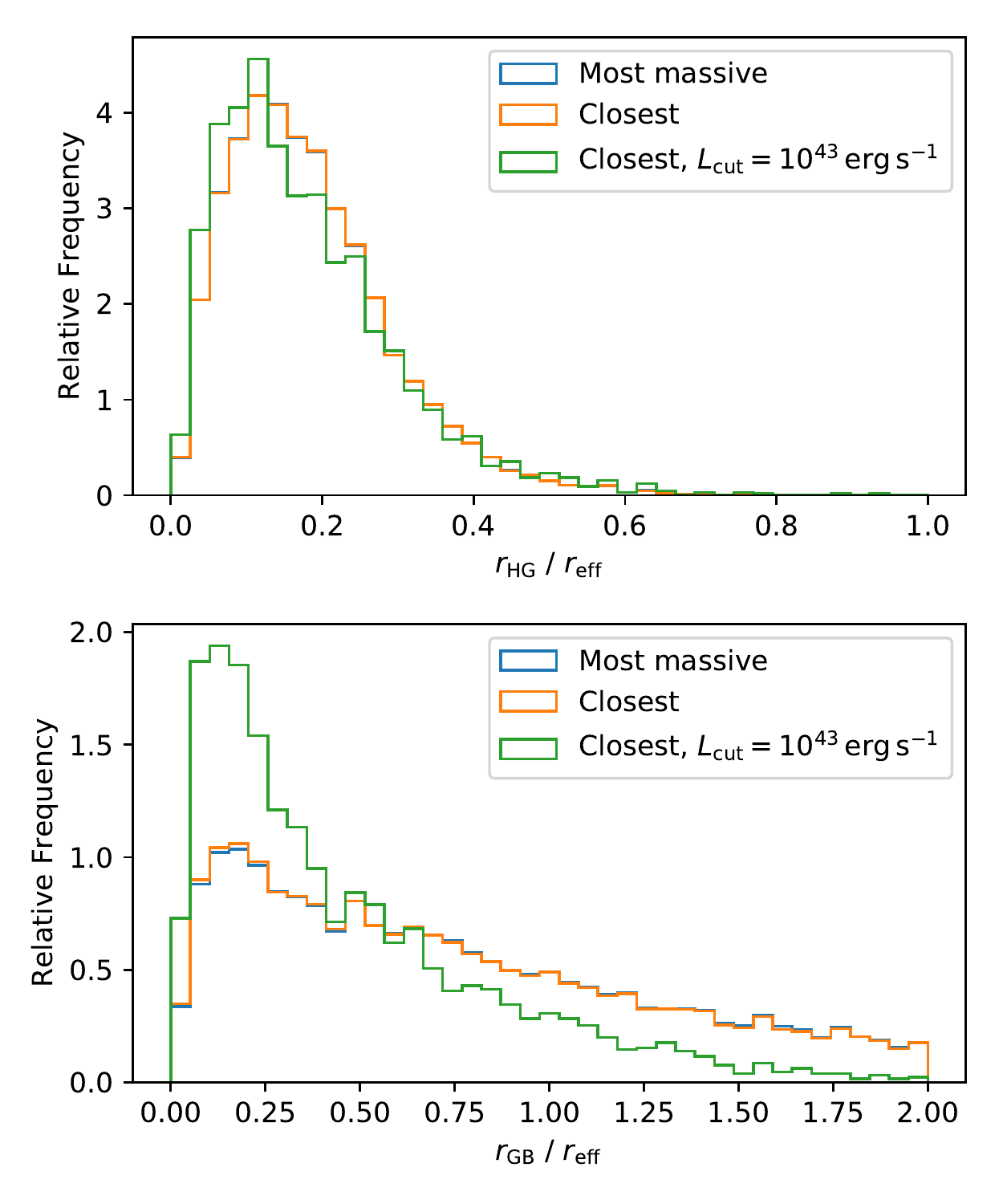}
	\caption{\label{fig:distance_ratio_plots} Distribution of halo-galaxy, $r_{\rm HG}$, and galaxy-BH, $r_{\rm GB}$, offsets in Horizon-AGN at $z=0.1$ as a multiple of the galaxy's effective radius, $r_{\rm eff}$, for systems obeying one of three selection cuts. In one case we select the most massive BH and in the other cases we select the BH closest to the galactic centre. For the case with a luminosity cut, $L_{\rm cut}$, we reject all BHs with luminosity $L < L_{\rm cut}$ before making the assignment. The galactic and halo centres are always close, whereas we can have very different offset BHs if we do not introduce a sufficiently stringent luminosity cut. The values of the halo-BH offsets, $r_{\rm HB}$, are very similar to $r_{\rm GB}$ and are thus not plotted. There is little difference between choosing the most massive or closest BH as the central BH.}
\end{figure}

In \autoref{fig:distance_ratio_plots} we plot the halo-galaxy, $r_{\rm HG}$, and BH-galaxy, $r_{\rm GB}$, distances in terms of $r_{\rm eff}$ with and without a luminosity cut. It is clear that for $L_{\rm cut} = 10^{43} {\rm \, erg \, s^{-1}}$, we preferentially select BHs closer to the centre of the galaxy than without this cut. In particular, $r_{\rm GB}<r_{\rm eff}$ for most BHs and the sharp cut-off observed at $r_{\rm GB} = 2 r_{\rm eff}$ is less visible with this luminosity cut. We also see in \autoref{fig:distance_ratio_plots} that if we assign the most massive BH as the central one, the distribution is almost identical to if we selected the closest BH without a luminosity cut. 99.3 per cent of galaxies which have a central BH using the `closest' method have the same central BH in the `most massive' method. If use $L_{\rm cut} = 10^{38} {\rm \, erg \, s^{-1}}$, the distribution is almost identical to those without a luminosity cut.

It may seem paradoxical that the `closest' method without a luminosity cut is more likely to select BHs that are further out than when we impose a non-zero $L_{\rm cut}$ (which does the same thing, however only after removing low-luminosity BHs) and is comparable to the `mass' method, which only uses BH mass. The reason is to do with which galaxies are retained in the sample. With the `closest' method all galaxies with BHs are retained, while the `$L_\text{cut}$' method only keeps galaxies with particularly luminous BHs. Since more luminous BHs tend to lie closer to their host galaxies' centres, the $L_\text{cut}$ method preferentially selects galaxies with small offsets. While the `closest' method will select the same BH in those galaxies (or one even closer to the centre), for the remaining galaxies with less luminous BHs the offsets are larger.

We therefore see that simply using the selection criteria of \cite{Volonteri_2016} produces a large tail of low luminosity BHs in the galaxy-BH offset distribution and that a cut is necessary to preferentially select the BHs near the centre of the galaxy. We use a range of selection criteria in this work. Our quasar and radio mode samples (see \autoref{sec:Quasar vs radio mode}) select the closest BH to the centre of the galaxy and impose cuts on accretion efficiency and luminosity. We detail the selection cuts used to compare to observational data in \autoref{sec:Observational Data}.

\section{Observational Data \& simulation selection cuts}
\label{sec:Observational Data}

\begin{figure*}
	 \includegraphics[width=\textwidth]{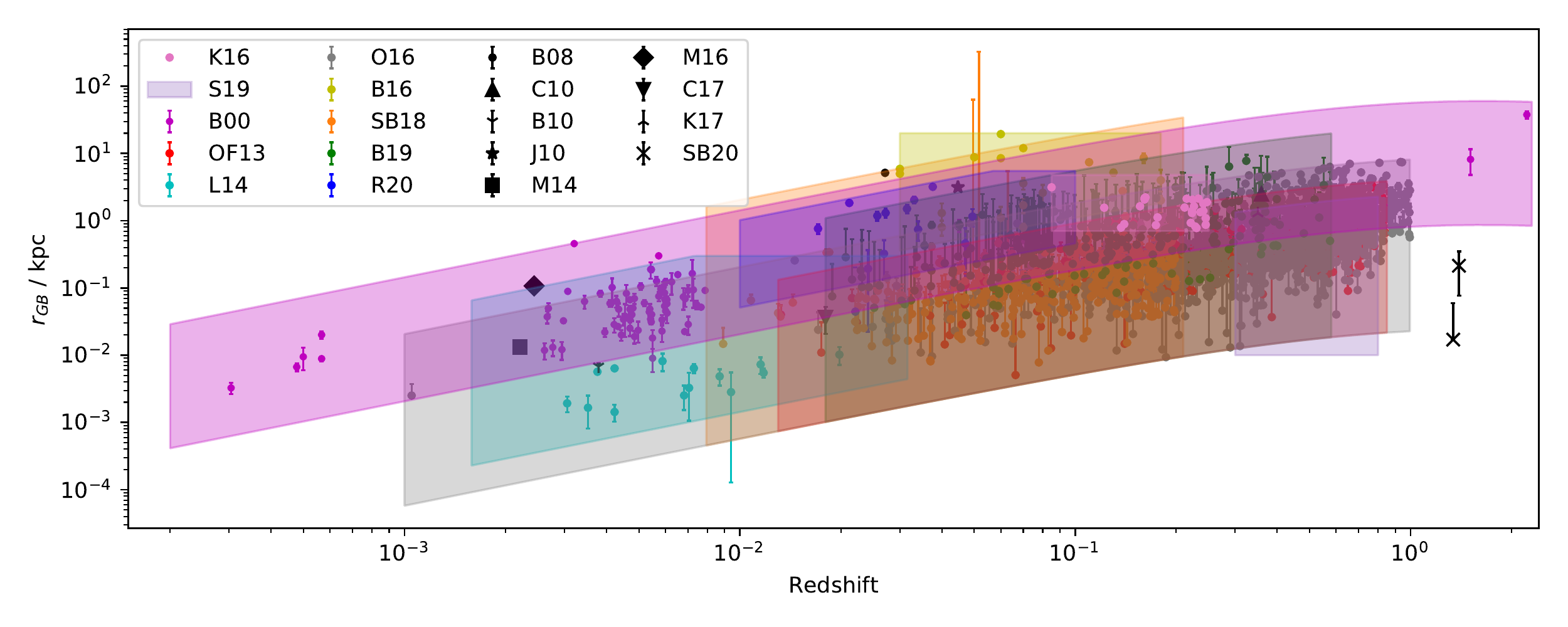}
	 \caption{Offset as a function of redshift for various observational datasets. The shaded regions indicate the approximate the regions each dataset is sensitive to, based on resolution and cuts made to the data. The OF13, SB18 and B19 data only show upper errorbars since the lower errorbars are of the same magnitude, but the logarithmic $y$ axis scale makes the plot confusing if the lower errors are also included. The lower error bar for one of the SB20 points has also been omitted as the offset is consistent with zero and this cannot be displayed with a logarithmic $y$ scale. The S19 region does not contain any points here, although in reality there are 8210 measurements (not publicly available). Black points correspond to studies of only one or two systems.}
	 \label{fig:offset_data}
\end{figure*}

Offset BHs have been observed through an array of techniques up to redshift $\sim 1.5$. In \autoref{fig:offset_data} we plot the BH displacement with respect to the centre of the host galaxy (e.g. the brightest part of the optical emission) projected onto the plane of the sky as a function of redshift for a number of observational samples. In this section we describe the various datasets and the region of the $r_{\rm GB}-z$ plane they are sensitive too, and these are summarised in \autoref{tab:data_summary}. Where appropriate, we also detail the selection cuts made to the Horizon-AGN sample in order to mimic those observations, as summarised in \autoref{tab:selection criteria}. We find that our AGN cut, where we only select BHs with $\chi > 0.01$, makes the samples almost identical, with each containing $\leq 533$ systems. The remaining cuts for specific samples remove up to an additional few tens of systems.

We note that offsets between the point identified as the BH and the galactic centre could be `true', where systems with large true offsets are likely to be mergers \citep{Barrows_2016}. However, misassociation, extended sources, double or lensed quasars, statistical outliers due to an extended tail \citep{Mingard_2018}, or the presence of a jet \citep{Kovalev_2017,Petrov_2017b} could also be responsible. Even smooth elliptical dust lanes could produce an apparent offset. Any values of the observed fraction of offset BHs should therefore be considered as upper limits for the intrinsic offset fraction, and this necessitates statistical methods to disentangle true vs merely apparent offsets.

\begin{table*}
    \caption{Summary of observational datasets used in this work. The number of objects may be smaller than given in the reference due to additional selection cuts made here. The resolutions are approximate, and are a combination of quoted, mean and fitted values. We provide the astrometric precision in the source plane for the SB20 sample, hence the small value. We do not give a resolution for S19, since these data are upper limits in the range $\sim5-20 {\rm \, kpc}$, nor for K16, since these are velocity offsets that are only approximately converted to physical offsets. The final columns indicates the wavelength used to determine the position of the BH.}
    \label{tab:data_summary}
    \centering
    \begin{tabular}{*6c}
        Name & Reference & Objects & Resolution $\left( \arcsec \right)$ & Redshift & BH Position \\
        \hline
        B00 & \citet{Binggeli_2000} & 78 & 0.42 & $z \sim 5 \times 10^{-3}$ & Optical \\
        OF13 & \citet{Orosz_2013} & 233 & 0.05 & $0.01 < z < 0.85$ & Radio \\
        L14 & \citet{Lena_2014} & 14 & 0.01 & $z < 0.02$ & Optical \\
        K16 & \citet{Kim_2016} & 26 & - & $z < 0.25$ & H$\alpha$ (Broad) \\
        O16 & \citet{Orosz_2016} & 1327 & 0.07 & $ z < 1$ & Radio \\
        B16 & \citet{Barrows_2016} & 48 & 0.06 & $z < 0.2$ & X-ray \\
        SB18 & \citet{Skipper_2018} & 345 & 0.06 & $z < 0.2$ & Radio \\
        B19 & \citet{Barrows_2019} & 254 & 0.18 & $z < 0.58$ & X-ray \\
        S19 & \citet{Shen_2019}& 8210 & - & $0.3 < z < 0.8$ & Optical \\
        R20 & \citet{Reines_2020} & 13 & 0.25 & $z < 0.055$ & Radio \\
        SB20 & \citet{Spingola_2020} & 2 & $5 \times 10^{-6}$ & $z \sim 1.4$ & Radio \\
    \end{tabular}
\end{table*}

\begin{table}
	\centering
	\caption{\label{tab:selection criteria} Selection criteria of the observational datasets that we use. We apply these also to Horizon-AGN to allow a fairer comparison. $L_{\rm R}$ and $L_{\rm X}$ are the minimum allowed radio and X-ray luminosities, respectively.}
	\begin{tabular}{*5c}
	 {} & {} &  \multicolumn{3}{c}{Minimum luminosity $ / \, {\rm erg \, s^{-1}}$} \\
	\hline
	{} & AGN cut & $L_{\rm R} \left( 5 {\rm \, GHz} \right)$ & $L_{\rm R} \left( 8.46 {\rm \, GHz} \right)$ & $L_{\rm X}$  \\
	\hline
	OF13 & \checkmark & & & \\
	L14 & \checkmark & $1.0 \times 10^{36}$ & & \\
	O16 & \checkmark & & & \\
	B16 & \checkmark & & & $1.0 \times 10^{42}$ \\
	SB18 & \checkmark & & $1.4 \times 10^{38}$ & \\
	B19 & \checkmark & & & $1.0 \times 10^{42}$ \\
	R20 & \checkmark & & & \\
	\end{tabular}
\end{table}

\subsection{Binggeli et al. 2000 (B00)}
\defcitealias{Binggeli_2000}{B00}

\citet{Binggeli_2000} investigated a sample of 78 `nucleated' dwarf galaxies (dE,N) in the Virgo cluster using a previous photometric study \citep{Binggeli_1991,Binggeli_1993}. They compare the position of the nucleus (brightest pixel) to the optical centre of the galaxy. The observed relation between the nuclear magnitude and ellipticity is predicted to be due to a central massive compact object (i.e. a BH). Objects in their sample are required to have an apparent \textit{B}-band magnitude brighter than $18 {\rm \, mag}$ and those which have bright stars near the centre are removed. The lower limit in \autoref{fig:offset_data} is set by the typical standard deviation of mean off-centre distance of $~0 \farcs 1$ and the upper limit is the maximum offset in the sample.

Since \citeauthor{Binggeli_2000} do not directly determine whether a BH resides in their nucleus, and there are no obvious selection cuts on BH luminosity, mass or accretion rate based on these observations, we do not produce a Horizon-AGN sample designed to mimic these data.

\subsection{Orosz \& Frey 2013 (OF13)}
\defcitealias{Orosz_2013}{OF13}

\citet{Orosz_2013} find the optical counterparts of 1297 radio sources from the International Celestial Reference Frame (ICRF2) in SDSS DR9 \citep{Ahn_2012}, of which 233 are classified by SDSS DR9 as extended (i.e. galaxy-AGN offsets) and the remainder are spatially unresolved and therefore appear as point-like quasars. The AGN position is determined from ICRF2 due to its superior astrometric precision as these are Very Long Baseline Interferometry (VLBI) observations, with the optical counterpart from SDSS required to lie within $0\farcs 5$, which is the upper limit in \autoref{fig:offset_data}. The sample covers the redshift range $0.01 < z < 0.85$. The cut-off in the published data of $3 \sigma \sim 0\farcs 17$ sets the lower limit in \autoref{fig:offset_data}.

As we are only interested in galaxy-AGN offsets so we exclude the quasar-like SDSS sources in this dataset. We also cut the Horizon-AGN systems so that $\chi > 0.01$.

\subsection{Lena et al. 2014 (L14)}
\label{sec:L14}
\defcitealias{Lena_2014}{L14}

\citet{Lena_2014} analyse archival \textit{Hubble Space Telescope} (\textit{HST}) images of 14 nearby ($d < 100 {\rm \, Mpc}$) bright elliptical galaxies containing low luminosity AGN. The selected galaxies were required to have an optically bright central point-like source and be free of heavy nuclear obscuration or other photometric irregularities. The offset is measured as the displacement between the photocentre (flux-weighted average of the centres of elliptical isophotes) and the AGN point source, which is modelled as a Gaussian. The lower limit of \autoref{fig:offset_data} is set by the \textit{HST} resolution and the upper limit is due to the search region of $2 \, \arcsec$.

Since the selection process requires a point-like AGN source, we only select quasar mode BHs from the simulation ($\chi > 0.01$) when mimicking this sample in Horizon-AGN. The minimum detected redshift from \cite{Lena_2014} is $z_{\rm min} = 3.0 \times 10^{-3}$, which corresponds to a luminosity distance of $d_{\rm L} = 13 {\rm \, Mpc}$. Converting the minimum detected flux (at $5 {\rm \, GHz}$) of $F_\nu = 1{\rm \, mJy}$ \citep{Capetti_2005} to a luminosity gives
\begin{equation}
	L_{\rm min} = 4 \pi d_L^2 F_\nu = 2.0 \times 10^{26} {\rm \, erg \, s^{-1} \, Hz^{-1}}.
\end{equation}
This must be converted to $L_{\rm R} =\nu  L_\nu$, thus we remove all systems from Horizon-AGN with $L_{\rm R} \left(5 {\rm \, GHz}\right) <  2.0 \times 10^{28} {\rm \, erg \, s^{-1} \, Hz^{-1}} \times 5 {\rm \, GHz} = 1.0 \times 10^{38} {\rm \, erg \, s^{-1}} $. To calculate the radio luminosity, we first calculate the X-ray luminosity, $L_{\rm X}$, from the bolometric luminosity, $L$, using \citep{Elvis_1994,Hopkins_2006}
\begin{equation}
	\label{eq:X-ray luminosity}
	L = 35.0 L_{\rm X}.
\end{equation}
Utilising the best-fitting fundamental plane of BHs \citep{Merloni_2003} we obtain the radio luminosity at $5 {\rm \, GHz}$
\begin{equation}
\label{eq:BH fundamental plane}
	\log \left( \frac{L_{\rm R} \left(5 {\rm \, GHz}\right) }{{\rm erg \, s^{-1}}} \right) = \xi_{\rm RX} \log \left( \frac{L_{\rm X}}{{\rm erg \, s^{-1}}} \right)
	+ \xi_{\rm RM}  \log \left( \frac{M_{\rm BH}}{\rm M_{\sun}} \right) + b_{\rm R},
\end{equation}
where $\xi_{\rm RX} = 0.60 \pm 0.11$, $\xi_{\rm RM} = 0.78^{+0.11}_{-0.09}$ and $b_{\rm R} = 7.33^{+4.05}_{-4.07}$. It should be noted that radio-loud and radio-quiet AGN obey different fundamental plane relations \citep{Wang_2006}. The value of $\xi_{\rm RX}$ above is less than the values reported in \citet{Wang_2006} of 1.39 for radio-loud and 0.85 for radio-quiet AGN. Our estimates of $L_{\rm R} \left(5 {\rm \, GHz}\right)$ are therefore smaller than if we used these alternative values of $\xi_{\rm RX}$, and we thus reject more systems here than in these alternative cases. We will find in \autoref{tab:Pcrit} that rejecting no systems based on luminosity (an extreme case of using larger $\xi_{\rm RX}$) has little effect on our results. Therefore, for simplicity we will apply the \citeauthor{Merloni_2003} relation for both radio-quiet and radio-loud AGN as this is the most stringent cut.

\subsection{Kim et al. 2016 (K16)}
\defcitealias{Kim_2016}{K16}

\citet{Kim_2016} looked for recoiling SMBHs from the $z < 0.25$ quasi-stellar objects in SDSS DR7 by targeting objects with broad lines that are blueshifted relative to the systemic velocity. They excluded those with highly asymmetric and widely separated broad-line velocity profiles (disc emitters) or those with double-peak emission lines (possible binary SMBHs). Performing a spectral decomposition of the H$\alpha$ and H$\beta$ lines, their final sample of 26 have kinematic offsets in H$\alpha$ of at least $69 {\rm \, km \, s^{-1}}$ (the instrument dispersion) and the H$\alpha$ and H$\beta$ velocities must agree within to 50 per cent. Assuming an AGN age of $\tau = 10 {\rm \, Myr}$ \citep{Blecha_2011}, the recoil velocity, $v_{\rm rel}$, can be converted to a physical offset as
\begin{equation}
    r_{\rm GB} = v_{\rm rel} \tau.
\end{equation}
We include these offsets in \autoref{fig:offset_data}, with the upper and lower limits set by the fibre radius and instrument dispersion respectively. We do not, however, use this data in the rest of the analysis since the spatial offsets are only estimates and depend on the chosen $\tau$. It should be noted that the magnitude of the inferred offsets appear consistent with the other data.

\subsection{Orosz et al. 2016 (O16)}
\defcitealias{Orosz_2016}{O16}
\label{sec:O16}

These data are currently unpublished, however were briefly presented by \citet{Orosz_2016}. The procedure for obtaining the offsets is identical to \citetalias{Orosz_2013}, however the matching is between SDSS DR12 \citep{Alam_2015} and mJIVE-20 \citep{Deller_2014}.
As before, we select objects classified by SDSS as galaxies, so obtain 1327 objects out of a total 2066. We note that the offset AGN candidates were not subject to follow-up observations to exclude spurious sources (e.g. jets, lensing systems etc.). It is therefore possible that this sample has a higher level of contamination than the other datasets. To quantify this, a Monte Carlo procedure similar to \citet{Orosz_2013} was performed (Orosz private communication) to determine the probability of false identification. The probability is twice as high in this sample than for \citetalias{Orosz_2013}, although both are $\sim 0.1$ per cent for a match radius of $500 {\rm \, mas}$, which is what we adopt.

\subsection{Barrows et al. 2016 (B16)}
\defcitealias{Barrows_2016}{B16}

\citet{Barrows_2016} searched for X-ray AGN by cross-matching sources from SDSS DR7 \citep{Abazajian_2009} with the $z<0.2$ OSSY catalogue \citep{Oh_2011}. These sources were then cross-correlated with \textit{Chandra} \citep{Evans_2010} and only sources containing \textit{i}- or \textit{z}-band SDSS images registered with \textit{Chandra} were kept. Sources with dust lanes or multiple emission peaks were removed, since these give false centroid positions. To reject stellar-mass objects, they required the difference between the observed luminosity in the range $2-10{\rm \, keV}$ and that expected from star formation, to be $> 3 \sigma$ and exceed $10^{42} {\rm \, erg \, s^{-1}}$. Further, the hardness ratio must be ${\rm HR} > -0.1$. This provides a sample of 48 type-II AGNs. The detection radius is less than $5 \, \arcmin$ from the observation aim point and the AGN must be within $20{\rm \, kpc}$ of centre of galaxy, which gives the upper limit in  \autoref{fig:offset_data}. The lower limit is given by the \textit{Chandra} resolution of $0\farcs 6$. We only plot the `offset sample' and not the full parent sample.

To select Horizon-AGN systems that are similar to these observations, we first cut on X-ray luminosity, $L_{\rm X}$, such that $L_{\rm X} > 10^{42} {\rm \, erg \, s^{-1}}$, where we calculate $L_{\rm X}$ using \autoref{eq:X-ray luminosity}. We then choose the closest remaining BH to the centre of the galaxy to be the central BH. The hardness ratio cut is designed to select only AGNs, so we use the quasar mode criterion $\chi > 0.01$.

\subsection{Skipper \& Browne 2018 (SB18)}
\label{sec:Skipper18}
\defcitealias{Skipper_2018}{SB18}

\citet{Skipper_2018} adopted a similar selection method to \citeauthor{Barrows_2016} for radio AGNs. They cross-matched sources from SDSS with the $z<0.2$ OSSY catalogue and compared\footnote{\url{http://www.jb.man.ac.uk/research/gravlens/class/class.html}} this with the Cosmic Lens All-Sky Survey (CLASS) \citep{Myers_2003,Browne_2003} to find sources within $10 \, \arcsec$, giving the upper limit of offsets in \autoref{fig:offset_data}. CLASS is a radio survey with the VLA and the images have an angular resolution of $200-250{\rm \, mas}$ at an observing frequency of $8.46 {\rm \, GHz}$. We plot the lower limit as twice the size of the point spread function (PSF) of SDSS as offsets smaller than this would unlikely be resolved. Matches outside the visible bulge of the galaxy were removed and so are galaxies in the starburst region of the BPT diagram \citep{Baldwin_1981}, which compares the flux ratios $F_{\ion{O}{iii}} \lambda 5007 / F_{\rm H\alpha}$ and $F_{\ion{N}{ii}} \lambda 6583 / F_{\rm H\alpha}$. Those known to be starburst galaxies and those missing optical emission line data products in the OSSY database were also removed.
This results in a final sample of 345 radio selected systems.

To mimic this selection process, we start by selecting the closest BH to the centre of each galaxy, as this is how they select their central BH. We do not impose a minimum galaxy mass cut at this stage but keep our BH mass cut. We reject all systems with $\chi \leq 0.01$. Finally, \citeauthor{Skipper_2018} only select candidates from the CLASS catalogue which have a flux density of at least $8 {\rm \, mJy}$ at $8.46 {\rm\,  GHz}$. We must therefore select AGNs that obey this criterion. We are not given the radio luminosities of the BHs in Horizon-AGN, so we do the following to make the correct cut. From the X-ray luminosity, $L_{\rm X}$, (\autoref{eq:X-ray luminosity}), we find the radio luminosity at $5 {\rm \, GHz}$ with \autoref{eq:BH fundamental plane}. The radio luminosity at $8.46 {\rm GHz}$ is
\begin{equation}
	L_{\rm R} \left(8.46 {\rm \, GHz}\right) = L_{\rm R} \left(5 {\rm \, GHz}\right) \left(\frac{\nu}{5 {\rm \, GHz}} \right)^{-\alpha_{\rm R}},
\end{equation}
for some spectral index $\alpha_{\rm R}$. We choose $\alpha_{\rm R}=0.4$ as this corresponds to the crossover between flat and steep spectra \citep{Merloni_2003}. Since the cutoff for CLASS is given as a flux, we wish to convert this to a minimum detectable luminosity. The minimum detected redshift is $z_{\rm min} = 8.9 \times 10^{-3}$, which corresponds to a luminosity distance of $d_{\rm L} = 38 {\rm \, Mpc}$. Converting the minimum detected flux of $F_\nu = 8{\rm \, mJy}$ to a luminosity gives
\begin{equation}
	L_{\rm min} = 4 \pi d_L^2 F_\nu = 1.4 \times 10^{28} {\rm \, erg \, s^{-1} \, Hz^{-1}}.
\end{equation}
Converting to $L_{\rm R} =\nu  L_\nu$, we remove all systems from Horizon-AGN with $L_{\rm R} \left(8.46 {\rm \, GHz}\right) <  1.4 \times 10^{28} {\rm \, erg \, s^{-1} \, Hz^{-1}} \times 8.46 {\rm \, GHz} = 1.2 \times 10^{38} {\rm ergs^{-1}} $.

The AGN cut reduces the number of systems to 533 and the luminosity cut, using $\xi_{\rm RX}$, $\xi_{\rm RM}$ and $b_{\rm R}$ from \autoref{sec:L14}, further reduces this to 530. Note that if we chose $\alpha_{\rm R} = 0$ (i.e. a completely flat spectrum) then none of the 533 systems would be removed. Further, if we use $\xi_{\rm RX}=0.76$, $\xi_{\rm RM}=0.71$ and $b_{\rm R}=1.31$, as found by \citeauthor{Merloni_2003} for flat-spectrum sources, we would not remove any system.

We also note that the flat radio spectra indicate that the emission from these sources is likely to be beamed. Our assumption of isotropic emission would then result in $L_{\rm min}$ being over-predicted, making our luminosity cut too severe. As in \autoref{sec:L14}, we choose the most stringent selection cuts and again note that we will find that our results would not change significantly if we did not make any luminosity cuts. Therefore, for simplicity, we assume isotropic emission in the luminosity cut and note than any anisotropy would not cause significant bias.

\subsection{Barrows et al. 2019 (B19)}
\defcitealias{Barrows_2019}{B19}

\citet{Barrows_2019} used an almost identical selection procedure to \citetalias{Barrows_2016} in a search for hyper-luminous X-ray sources by again comparing SDSS and \textit{Chandra} data. In the 2019 sample, the search radius is changed from within the $3\arcsec$ fibre to within two Petrosian radii. Further, the latest sample imposes a much stricter cut on the compactness of the source and the errors, specifically the X-ray source errors, are estimated differently. Although some galaxies are common to this sample and \citetalias{Barrows_2016} we plot them separately in \autoref{fig:offset_data} as the offset may differ between the two. To compare to Horizon-AGN we make the same selection cuts as for \citetalias{Barrows_2016}.

\subsection{Shen et al. 2019 (S19)}
\defcitealias{Shen_2019}{S19}

\citet{Shen_2019}, using the technique of Vastrometry \citep{Hwang_2020}, were able to put upper limits on the magnitude of the offsets from 8210 AGNs in \textit{Gaia} DR2 \citep{Gaia_2016,Gaia_2018}. This technique utilises the astrometric jitter caused by stochastic variability of the AGN to place upper limits on the BH displacement. \textit{Gaia} cannot resolve separations $\la 1 {\rm \, kpc}$ in the desired redshift range of $0.3 < z < 0.8$, but this method enables upper limits to be determined in the range $\sim 5 {\rm \, pc}$ to $\sim 1 {\rm \, kpc}$, as indicated in \autoref{fig:offset_data}.

Since this method does not measure the projected off-nucleus distance, we do not compare \citeauthor{Shen_2019}'s results directly to Horizon-AGN, but we do note that they find that AGNs are well centred at these redshifts, with 99, 90 and 40 per cent of AGNs within $1 {\rm \, kpc}$, $500 {\rm \, pc}$ and $100 {\rm \, pc}$ respectively.

\subsection{Reines et al. 2020 (R20)}
\defcitealias{Reines_2020}{R20}

\citet{Reines_2020} cross-correlated galaxies from the NASA-Sloan Atlas\footnote{\url{http://www.nsatlas.org/documentation}} ($z < 0.055$) with a sub-sample of dwarf galaxies from the Very Large Array (VLA) Faint Images of the Radio Sky and Twenty-centimetres (FIRST) Survey \citep{Becker_1995}, requiring a match radius $\leq 5 \, \arcsec$, which is approximately the resolution of the $1.4 {\rm \, GHz}$ radio observations. This gives 186 matches, after enforcing a maximum stellar mass of $M_{\star} \leq 3 \times 10^9 {\rm \, M_{\sun}}$ and absolute magnitude cuts of $M_{g}, M_{r} > -20$ to prevent spurious mass estimates. Sources clearly not dwarf galaxies were removed and 111 objects were observed with the VLA at higher frequencies ($\sim 8-12{\rm \, GHz}$) and therefore higher resolution. The match radius and the typical angular resolution of the VLA follow-up observations of $~0.25 \, \arcsec$ define the shaded region in \autoref{fig:offset_data}. Only those with $> 3 \sigma$ radio detections were retained, and those with optical counterparts which appear to come from unrelated background point sources were removed. Finally, the VLA detections were used to eliminate samples with emission from thermal \ion{H}{ii} regions, individual supernova remnants or young supernovae. This results in a sample of 13 compact radio sources, which are almost certainly AGNs.

The VLA analysis ensures that only AGNs are selected, hence we cut the Horizon-AGN data such that $\chi>0.01$. The minimum mass of these galaxies is $6.5 \times 10^{9} {\rm \, M_{\sun}}$, which is small compared to typical masses in Horizon-AGN. Since this cut was only introduced to ensure reasonable mass estimates we do not introduce a mass cut here. We explore the impact of galaxy mass on the BH offsets in \autoref{sec:Effect of galaxy mass}.

\subsection{Spingola \& Barnacka 2020 (SB20)}
\defcitealias{Spingola_2020}{SB20}

Recent work by \citet{Spingola_2020} demonstrates how to exploit the non-linear magnification of gravitational lensing to access parsec scales at cosmological redshifts when an AGN lies close to a caustic curve. Using VLBI radio observations, they detected one object ($z = 1.34$) with optical and radio centres within $\sim 40 {\rm pc}$ of each other, and another ($z = 1.39$) with a radio-optical offset of $214 \pm 137 {\rm \, pc}$.

Currently this sample size is too small for comparison with Horizon-AGN. Nonetheless, the astrometric precision achieved is impressive. This methodology can be applied to future optical and radio surveys of gravitational lensing systems, which will allow to detect offset BHs at high redshifts.

\subsection{Other observations}

The remaining observations in \autoref{fig:offset_data} are of single systems, hence we cannot make statistical comparisons of these to Horizon-AGN. We include them for visual comparison with the other data, and summarise them here.

\begin{itemize}
    \item \textbf{B08} \citep{Barth_2008} is the offset Seyfert-2 nucleus in the system NGC 3341. This system consists of three nuclei, but the optical emission lines suggest only one of these harbours an AGN, however this is not the primary nucleus. Further optical, radio and X-ray observations \citep{Bianchi_2013} support this conclusion.
    \item \textbf{C10} \citep{Civano_2010} is one of the best studied offset AGNs, CXOC J100043.1+020637 (CID-42), with optical, high resolution X-ray \citep{Civano_2012} and radio \citep{Wrobel_2014,Novak_2015} observations consistent with an AGN ejected from a separate compact region. Simulations \citep{Blecha_2013} also support this conclusion, as opposed to a dual-AGN system. The offset used in \autoref{fig:offset_data} is the measured displacement between the two compact optical sources in the \textit{HST}/ACS image.
	\item \textbf{B10} \citep{Batcheldor_2010} is the displaced SMBH in M87*. The offset is measured using archival \textit{HST} data, with the offset between the photocentre of the galaxy and AGN point source.
	\item \textbf{J10} \citep{Jonker_2010} is the source CXO J122518.6+144545: an X-ray source from \textit{Chandra} offset from a galaxy from SDSS DR7. It is unknown whether the source is a recoiling SMBH, bright ultra-luminous X-ray source with a bright optical counterpart, or a very blue Type IIn supernova. A candidate optical counterpart to the X-ray source is found in archival \textit{HST} data.
	\item \textbf{M14} \citep{Menezes_2014} is an AGN in NGC 3115 observed to be off-centred from the photometric centre, using the Gemini Multi-Object Spectrograph mounted on the Gemini-South telescope.
	\item \textbf{M16} \citep{Menezes_2016} is a Seyfert 2-like source offset from the central stellar cluster in NGC 3621, observed using the Gemini Multi-Object Spectrograph. An X-ray source found at the centre of the stellar cluster suggests this may not in fact be an offset BH.
	\item \textbf{C17} \citep{Condon_2017} is a SMBH in the cluster ZwCl 8193, which is offset from both the brightest cluster galaxy ($\sim 8.5 {\rm \, kpc}$) and its host galaxy, a small, optically faint radio galaxy ($\sim 0.1 {\rm \, kpc}$).
	\item \textbf{K17} \citep{Kim_2017} is the source CX0 J101527.2+625911, a recoiling or dual SMBH discovered from the \textit{Chandra} Source and SDSS Cross-Match Catalogue. No X-ray source is observed at the galaxy centre, supporting the recoiling SMBH hypothesis.
\end{itemize}

We note that many other catalogues are used to investigate of optical-radio offsets, such as \textit{Gaia} DR1 \citep{Lindegren_2016,Mingard_2016} and DR2 \citep{Lindegren_2018,Mingard_2018} combined with ICRF2, the Rio catalogue \citep{Assafin_2013} or the catalogue of \citet{Zacharias_2014}. Although some of the conclusions of these studies are similar to this work, the lack of photometry means we cannot guarantee that these offsets are the same as are of interest here, so we do not use them. In future work, we hope to be able to use upcoming \textit{Gaia} data releases, once photometric classifications are included, to expand our sample size.

\section{Comparing Horizon-AGN with the data}
\label{sec:Comparing Horizon-AGN with data}

In the previous section we described observational searches for offset AGN, which should be interpreted as finding `candidate' rather than necessarily `true' offset systems. Despite this limitation, we believe it is important to study the exact nature of any discrepancy between these and the simulations in order both to understand it fully and to improve the simulations. However, due to Horizon-AGN having a much coarser spatial resolution than the observations, we cannot simply compare the magnitudes of the offsets since the resolution makes the simulated values appear much larger. Therefore in this section we develop the statistical framework required to remove the effect of resolution and hence compare the observed and simulated offsets fairly.

\subsection{Accounting for finite resolution}
\label{sec:Accounting for finite resolution}

Since the simulations and observations have a finite resolution, we need a prescription to quantify the number of intrinsically offset BHs. Inspired by \citet{Skipper_2018}, we bin the $d$-dimensional offsets and fit to this two components:
\begin{equation}
\label{eq:Offset distribution fit}
	p_{d}\left( r \right) = a_d \left( r \right) + b_{d} \left( r \right).
\end{equation}
$p_d(r)$ is the probability of an offset $r$ (as described by the number of counts in the corresponding histogram bin), $a_d(r)$ is the probability distribution function for apparent offsets produced by the finite spatial resolution of the observation or simulation, and $b_d(r)$ is the probability distribution function for intrinsic offsets. Assuming a Gaussian distribution for each Cartesian component, $a_d$ takes the form
\begin{equation}
\label{eq:ad definition}
    a_d \left( r \right) = A r^{d - 1}  \exp\left( - \frac{r^2}{2\sigma^2} \right).
\end{equation}
\citeauthor{Skipper_2018} give an explicit model for $b_{d} \left( r\right)$, however we do not need to know this to find the fraction of BHs that are intrinsically offset. Instead we enforce normalisation, 
\begin{equation}
    \int_0^\infty b_{d} \left( r \right) {\rm d} r = 1 - \int_0^\infty a_{d} \left( r \right) {\rm d} r,
\end{equation}
so that
\begin{equation}
\label{eq:Offset fraction}
\begin{split}
	P_{d} \left( {\rm offset} \right) &= \frac{\int_0^\infty b_{d} \left( r \right) {\rm d} r}{\int_0^\infty \left(  a_{d} \left( r \right) +b_{d} \left( r \right) \right) {\rm d}r } \\
	& = 1 - 2^{d/2 -1} \Gamma \left( \frac{d}{2} \right) A \sigma^{d},
\end{split}
\end{equation}
where $\Gamma(x)$ is the gamma function. We fit for $A$ and $\sigma$ in the region $r_{\rm GB} < r_{\rm cutoff}$, for some cut-off offset $r_{\rm cutoff}$, with the expectation that $\sigma \sim 1\,{\rm kpc}$, the approximate spatial resolution of Horizon-AGN.
We find that the fitting parameters converge by $r_{\rm cutoff} \sim 3 {\rm \, kpc}$, but reducing this to $r_{\rm cutoff} \sim 2 {\rm \, kpc}$ does not significantly change the results. Choosing $r_{\rm cutoff} \gg 3 {\rm \, kpc}$ produces unrealistic fits, as this tries to make $a_d$ fit the tail too well, to the detriment of the small $r$ part of the distribution.

For each system, $i$, with $d$-dimensional offset $r_i$, we can now define the probability that the offset is intrinsic offset as
\begin{equation}
	w \left( r_i \right) = \frac{ b_d \left( r_i \right) }{ a_d \left( r_i \right) +b_{d} \left( r_i \right)}.
\end{equation}
Hence, we need to find the functional form of $b_{d}(r)$, since the normalisation condition is no longer sufficient. \citet{Skipper_2018} assume $b_2 \left( r \right)$ takes the form of a decaying exponential. We would like to have a form for $b_{d}$ such that its marginal distribution gives $b_{d-1}$. It is not obvious how to do this for an exponential, although the good fit obtained by \citeauthor{Skipper_2018} shows that a functional form that decays exponentially at large $r$ is desirable. We choose a generalised symmetric Laplace distribution \citep{Kotz_2001,Kozubowski_2013}, since all marginal distributions of a multivariate Laplace distribution are also multivariate Laplace distributions.

Working in coordinates centred on the galactic centre, we assume that the mean vector offset is zero. If we assume that the components of the displacements along each axis are independent and that there is no preferred direction, the generalised symmetric Laplace distribution for a displacement $r$ is
\begin{equation}
	g_d \left( r \right) = \frac{2^{1 - d/2}}{\Gamma \left( \frac{d}{2} \right)} \lambda  \left( \lambda r \right)^{d/2}  {\rm K}_{1 - d/2} \left( \lambda r \right),
\end{equation}
with ${\rm K}_n$ the modified Bessel function of the 2\textsuperscript{nd} kind.

This form could arise in a model where the true displacement follows a multivariate variance gamma process \citep{Madan_1990}, which is equivalent to a Brownian motion subordinated to a Gamma subordinator. The normally-distributed Brownian motion would correspond to the BH stochastically wandering in the local environment and the displacements due to mergers would be modelled as independent gamma distributed increments with a shape parameter of 1.

\begin{figure}
	\centering
	 \includegraphics[width=\columnwidth]{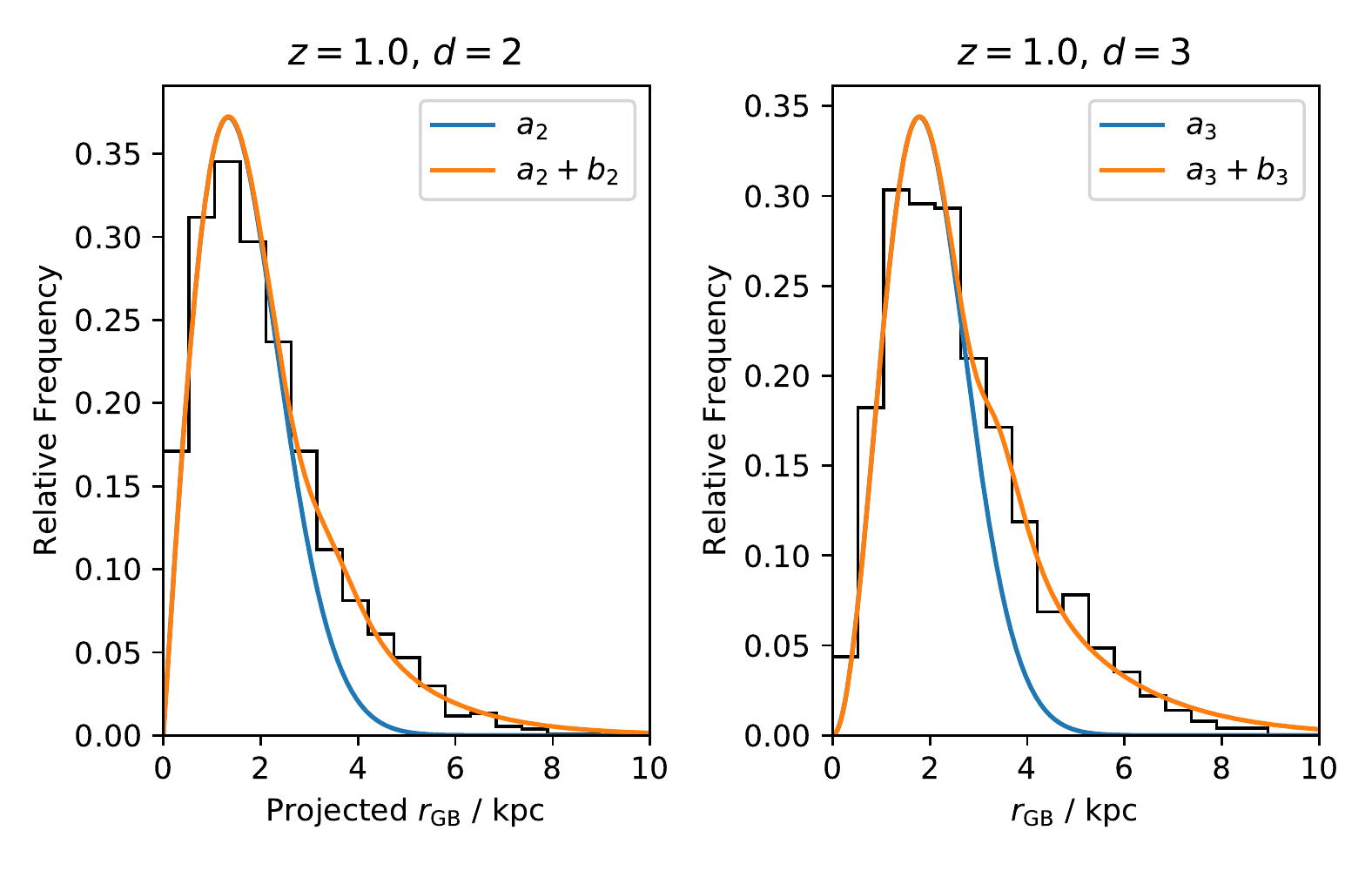}
	 \includegraphics[width=\columnwidth]{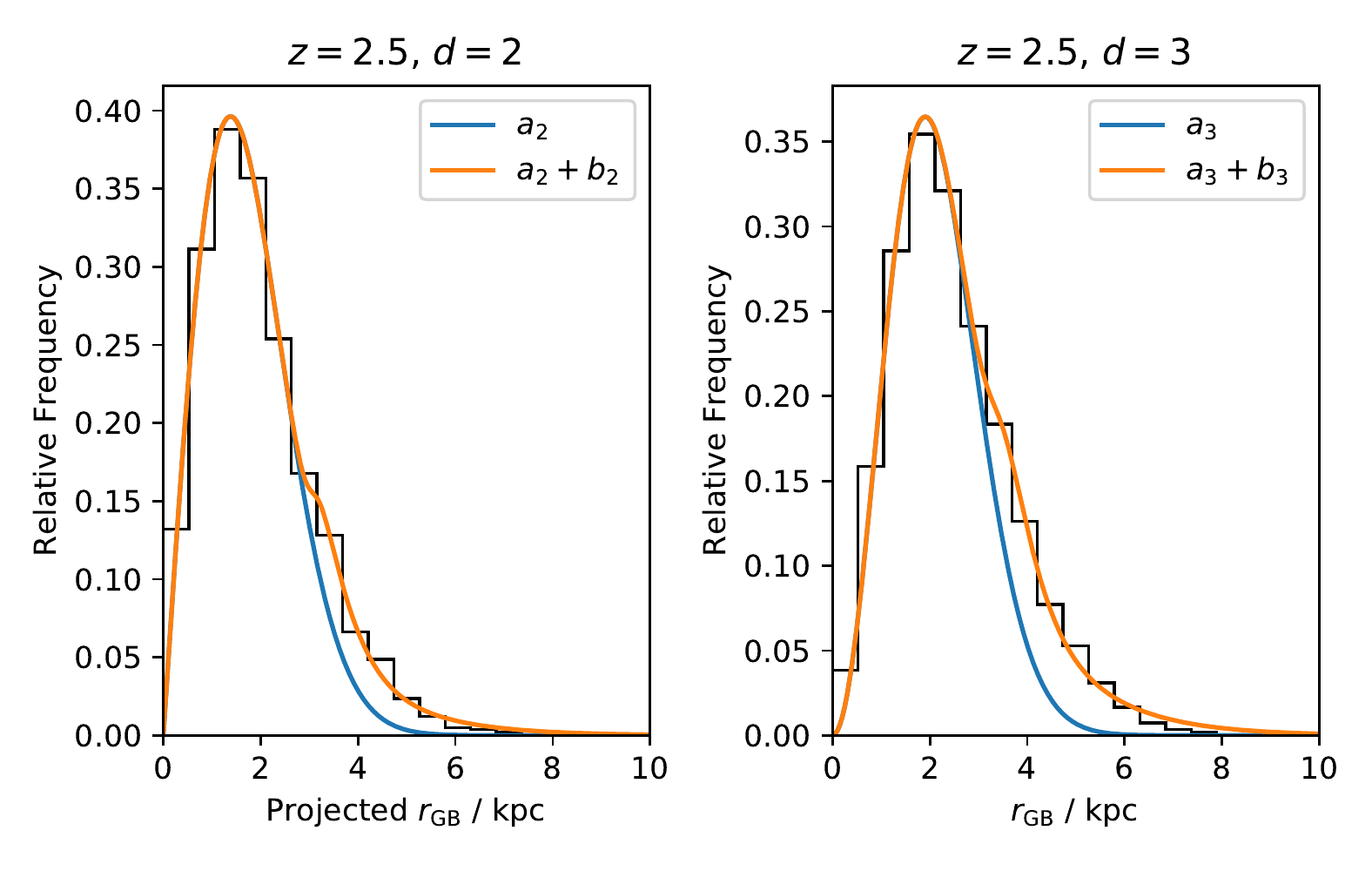}
	 \caption{Distributions of offsets for the quasar mode systems in Horizon-AGN and the fits for both the true ($d=3$) and projected ($d=2$) offsets. The $a_d$ term captures the offsets due to the finite resolution and $b_d$ models the intrinsically offset population. The $z=1.0$ sample has a higher fraction of intrinsically offset BHs and hence requires a larger contribution from $b_d$ than the $z=2.5$ sample.}
	 \label{fig:fitting_example}
\end{figure}

Unfortunately, with the resolution of Horizon-AGN, the peak of $g_d(r)$ occurs at a similar value to the peak of $a_d (r )$ for appropriate fitting parameters if we assume $b_d \propto g_d$. This suppresses the contribution from the resolution to very low levels and pushes $\sigma < 1 {\rm \, kpc}$. In order to suppress $b_d$ at small $r_{\rm GB}$, we multiply $g_d$ by a sigmoid function and thus arrive at
\begin{equation}
\label{eq:bd definition}
	b_d \left( r \right) = B r^{d / 2} {\rm K}_{1 - d/2} \left( \lambda r \right) \frac{1}{1 + \exp \left( - \kappa \left( r - \nu \right) \right)}.
\end{equation}
$\nu$ is an additional free parameter to be fitted, which we expect to be around 3 kpc because this is where $p_d(r)$ starts to deviate from $a_d(r)$ (\autoref{fig:fitting_example}). Although this removes the property that $b_{d-1}$ is the marginal of $b_{d}$, we anticipate that future higher-resolution simulations would not require the sigmoid function and thus this property would be conserved.

In \autoref{fig:fitting_example} we demonstrate these fits for a highly offset sample ($z=1.0$) and one with a low fraction of offset BHs ($z=2.5$) where we use quasar mode systems only. The results are similar for the radio mode, but we choose to only plot the quasar mode for clarity and because this is the mode used when comparing to observations. Note we assume that the shape parameter remains equal to 1 at all redshifts for simplicity. The degree to which a sample is intrinsically offset is determined by how well $a_d$ fits the distribution.

\subsection{Probability of consistency}

After making selection cuts to match the simulated galaxies to the observed ones, we wish to find the probability of generating a dataset with up to as many intrinsically offset BHs as is observed. To do this, using the parameter vector $\bm{\Omega} = \{ A, \sigma, B, \lambda, \kappa, \nu \}$ from the fit to \autoref{eq:Offset distribution fit}, we find $\{ w_i \}$. For each system, we then draw a random number $R_i \in [0, 1]$ from a uniform distribution $U[0,1]$ and define the random variable $a_i$ such that
\begin{equation}
	a_i = 
		\begin{cases}
			\mbox{$1$} & \text{if} \quad w_i > R_i \\
			\mbox{$0$} & \text{if} \quad w_i \leq R_i.
		\end{cases}
\end{equation}
For $\mathcal{N}$ systems, the fraction of offset BHs is
\begin{equation}
	f_{\rm offset} = \frac{\sum_i a_i}{\mathcal{N}}.
\end{equation}
Repeating this procedure $N_{\rm MC}=10^5$ times, we find a distribution of $f_{\rm offset}$, $p\left( f_{\rm offset} | \bm{\Omega}\right)$.

Using the means and errors on the fits for $\bm{\Omega}$, we draw $N_{\rm dist}$ random values of $\bm{\Omega}$ from a Gaussian distribution and repeat the above analysis for each of those samples. Since $\bm{\Omega_B} = \{B, \lambda, \kappa, \nu \}$ is determined after finding $\bm{\Omega_A} = \{ A, \sigma \}$, we draw $\bm{\Omega_A}$ from its multivariate Gaussian distribution first, fit the residuals to obtain means and covariances for $\bm{\Omega_B}$ and then draw $\bm{\Omega_B}$ from the resulting multivariate Gaussian distribution. We reject any of the iterations if $\omega < 0$ for any $\omega \in \bm{\Omega}$ since these are unphysical, or if the fit for $\bm{\Omega}_B$ is unsuccessful for the given $\bm{\Omega}_A$. We combine these $N_{\rm dist}$ samples to find the distributions of the parameters describing $p\left( f_{\rm offset} | \bm{\Omega}\right)$.

\begin{figure}
	 \includegraphics[width=\columnwidth]{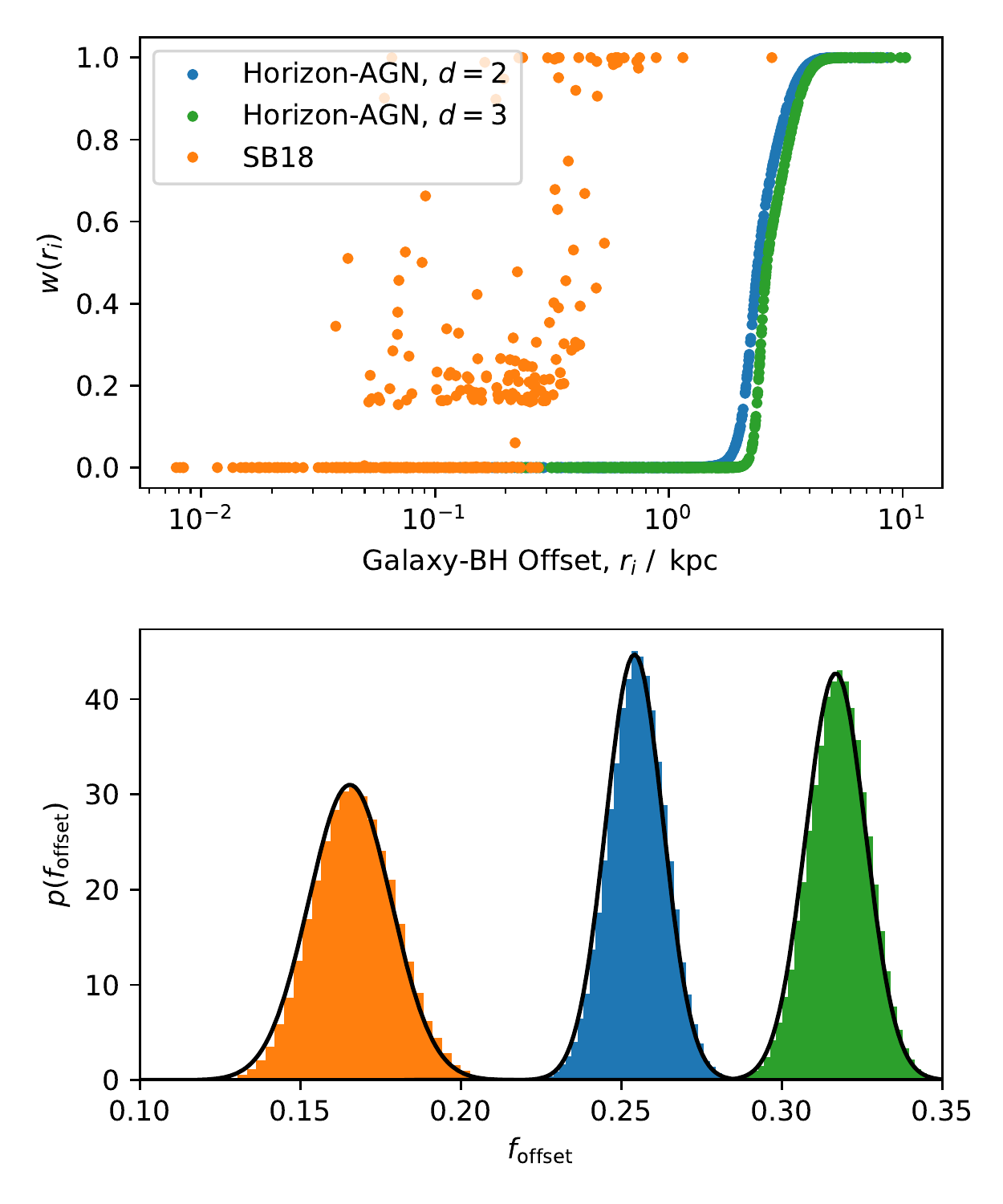}
	 \caption{\label{fig:compare_h_agn_to_skipper_test} \textit{Upper}: The probability of a BH being intrinsically offset, $w \left( r_i \right)$, as a function of the offset, $r_i$. The points correspond to individual systems. The relation is not monotonic for SB18, as $w$ is determined by the angular offset, whereas here we plot the offset distance. \textit{Lower}: The probability density functions of the fraction of intrinsically offset BHs, $p\left( f_{\rm offset} \right)$, using the best fit parameters. The black lines are Gaussians with the same mean and standard deviation as the data. For all these plots we use the selection procedures outlined in \autoref{sec:Skipper18}.}
\end{figure}

In \autoref{fig:compare_h_agn_to_skipper_test} we plot $w\left( r_i \right)$ and $p\left( f_{\rm offset} \right)$ for the best-fitting $\bm{\Omega}$ for the \citetalias{Skipper_2018} sample and the Horizon-AGN sample designed to mimic these observations (\autoref{sec:Skipper18}). We see that $p\left( f_{\rm offset} \right)$ is well approximated by a Gaussian, hence we will characterise the distributions by their mean, $\mu \left( \bm{\Omega} \right)$, and their standard deviation, $s \left( \bm{\Omega} \right)$.

\begin{figure*}
	\centering
	\includegraphics[width=\textwidth]{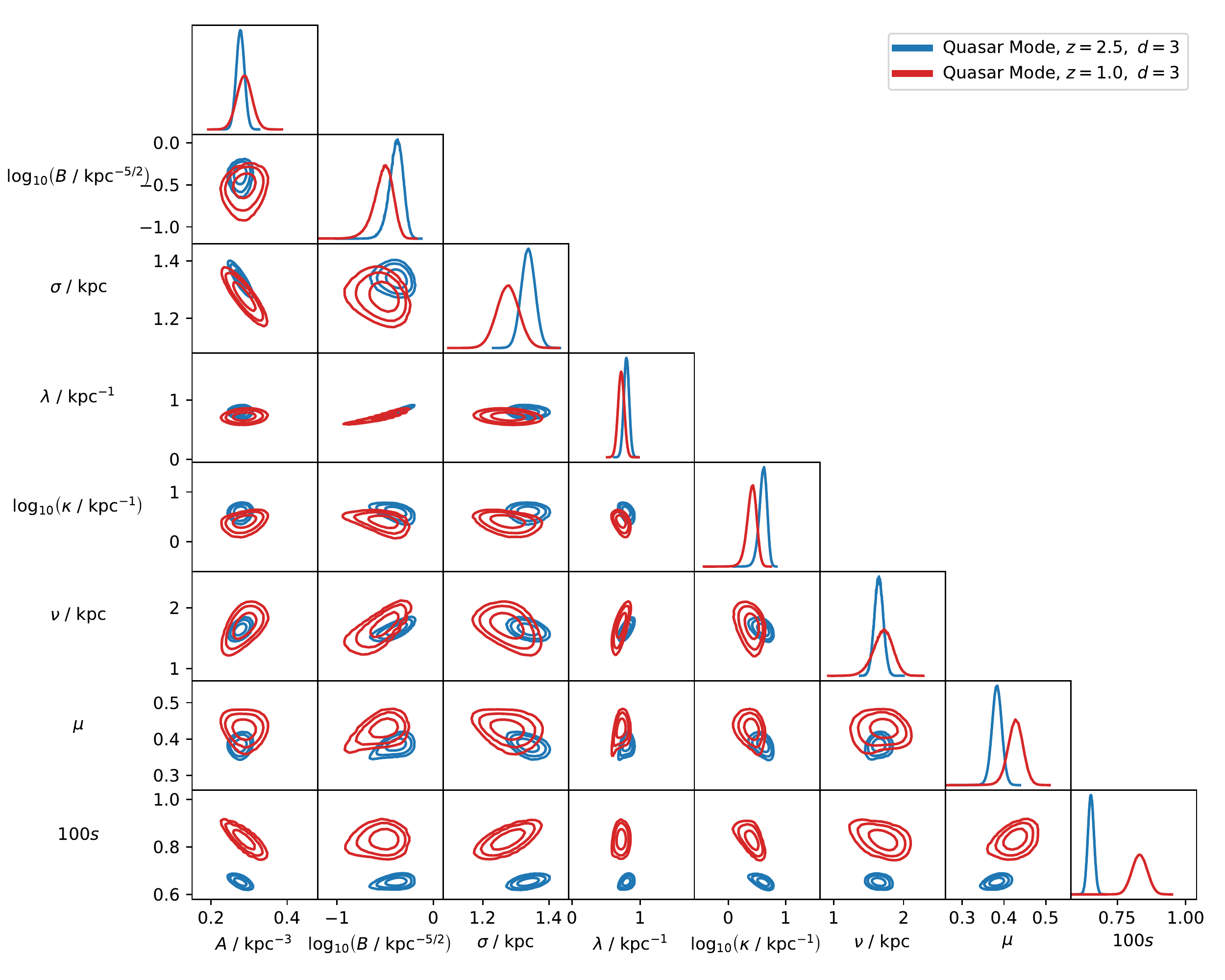}
	 \caption{\label{fig:f_offset_compare_h_agn_redshift_fits} The distributions of fitting parameters $A$, $B$, $\sigma$, $\lambda$, $\kappa$ and $\nu$ (\cref{eq:Offset distribution fit,eq:ad definition,eq:bd definition}) and the parameters describing the distribution of $f_{\rm offset}$: the mean, $\mu$ and standard deviation, $s$, where we consider the true offsets ($d=3$) of the quasar mode systems at the given redshifts. We fit the offsets to the sum of a Gaussian of width $\sigma$ and the product of a sigmoid function and generalised symmetric Laplace distribution, with typical scale $\lambda$. $\kappa$ and $\nu$ parametrise the steepness and position of the sigmoid respectively. $A$ controls the weight of the Gaussian term, and $B$ has the same role but for the non-Gaussian term (see \autoref{eq:ad definition} and \autoref{eq:bd definition}). The contours show the 1, 2 and $3\sigma$ confidence intervals.}
\end{figure*}

The results for the fits for the distributions in \autoref{fig:fitting_example} are plotted in \autoref{fig:f_offset_compare_h_agn_redshift_fits}. Since the $z=1.0$ sample is more offset than at $z=2.5$, we see that it has a larger value of $\mu$. We also note that $\sigma$ is comparable between the two fits, since the resolution of Horizon-AGN is constant with redshift at$\sim 1 {\rm \, kpc}$.

For each of the $N_{\rm dist}$ runs, we find the probability of generating a sample from Horizon-AGN with up to a fraction $f_{\rm crit}$ of offset BHs, which is
\begin{equation}
	\label{eq:Pcrit}
	\begin{split}
		P_{\rm crit} & \equiv P\left( f_{\rm offset} \leq f_{\rm crit} \right) \\
		&= \int_{- \infty}^{f_{\rm crit}} \frac{1}{\sqrt{2 \pi} s} \exp \left( - \frac{ \left( f - \mu \right)^2}{2 s^2} \right) {\rm d} f \\
		&= \frac{1}{2} \left[ 1 + \erf \left( \frac{f_{\rm crit} - \mu}{s \sqrt{2}} \right) \right],
	\end{split}
\end{equation}
since $p\left( f_{\rm offset} | \bm{\Omega}\right)$ is approximately Gaussian. We can then find the mean and error of $P_{\rm crit}$ from the $N_{\rm dist}$ iterations.

To prevent poor fits to the tail, we enforce $\nu \leq 4 {\rm \, kpc}$ (\autoref{eq:bd definition}), although our results are not sensitive to this choice; relaxing this to $\nu \leq 8 {\rm \, kpc}$ changes $\mu$ for the quasar-mode sample at $z=0$ with $d=3$ from $0.40\pm0.06$ to $0.36\pm0.12$. It is also necessary, for $d=2$, to demand $\nu \geq \nu_{\rm min}$, where we set $\nu_{\rm min} = 2.3 {\rm \, kpc}$. Once again, our results are not sensitive to this choice; if we remove this constraint for the quasar-mode sample at $z=0$ with $d=2$, then $\mu$ changes from $0.25\pm0.05$ to $0.26\pm0.05$.

\section{Results}
\label{sec:Results}

In this section we start by analysing the Horizon-AGN systems, by looking at the impact of the mode of feedback (\autoref{sec:Quasar vs radio mode}) and the correlations between offsets and the galaxy and halo properties (\autoref{sec:Correlation of offsets with galaxy and halo properties}). We then assess the compatibility of the observations with each other and with Horizon-AGN in \autoref{sec:Compatibility of observational datasets} and \autoref{sec:Comparison of Horizon-AGN with observations} respectively.

\subsection{Quasar vs radio mode}
\label{sec:Quasar vs radio mode}

\begin{figure}
	 \includegraphics[width=\columnwidth]{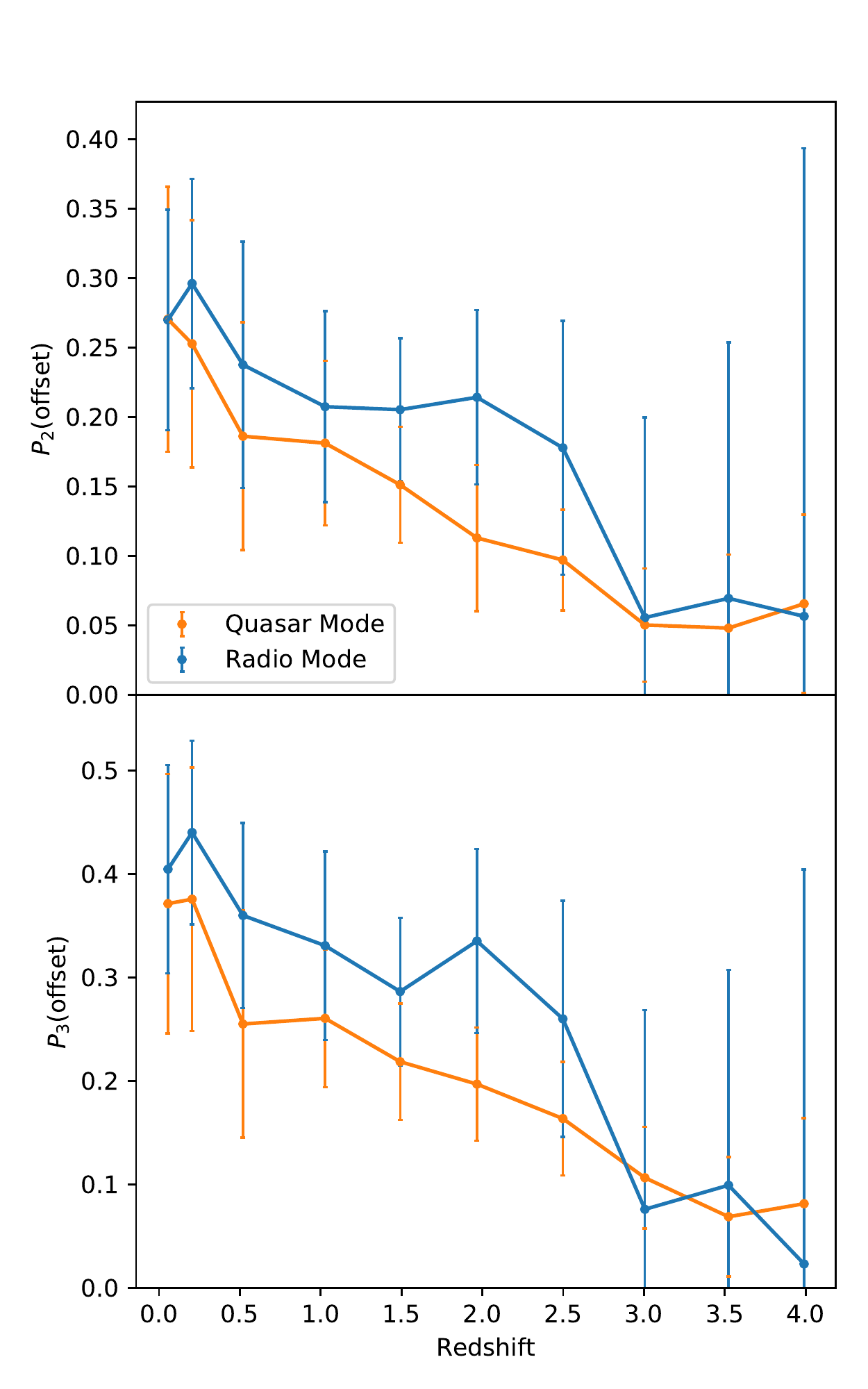}
	 \caption{Fraction of offset BHs (see \autoref{eq:Offset fraction}) as a function of redshift in Horizon-AGN. The BHs are selected as the closest BH to the centre of their galaxy. The `Quasar Mode' BHs have $\chi > 0.01$ and the `Radio Mode' BHs have the same luminosity range as the `Quasar Mode' BHs, but with $\chi \leq 0.01$. The lower panel uses the full 3D offset while the upper panel projects to 2D using an observer at the centre of the simulation box. We find Radio Mode BHs to be more intrinsically offset than Quasar Mode ones at $0<z<3$.}
	 \label{fig:h_agn_redshift_evolution}
\end{figure}

To compare the effect of selecting quasar or radio mode BHs at each redshift we start by assigning the closest BH to the centre of the galaxy as the central BH. At each redshift, we find the minimum luminosity of the quasar mode BHs and only consider BHs with luminosities greater than this. At $z=0.1$ this corresponds to a luminosity cut of $2.6 \times 10^{43} {\rm \, erg \, s^{-1}}$. We split the sample into two sets, those in the quasar mode ($\chi > 0.01$) and those in the radio mode ($\chi \leq 0.01$). We fit the distributions of offsets and plot the resulting $P_d ({\rm offset})$ for the two samples as a function of redshift in \autoref{fig:h_agn_redshift_evolution}.

At all redshifts, the radio mode quasars are more offset than those in the quasar mode. This is to be expected since by definition the quasar mode BHs operate at a higher fraction of the Eddington rate, so these will tend to lie closer to the centre of the galaxy where the gas density is higher. We note that this could also arise due to the different implementations of AGN feedback between the two samples. For $z \ga 3$ the two modes have comparable values of $P_d ({\rm offset})$, but this is also where the errors on the radio mode sample become large.

It is also clear in \autoref{fig:h_agn_redshift_evolution} that BHs are more central at earlier epochs. This is consistent with the picture that BHs initially reside at the centres of galaxies and then move off-centre later due to interactions with other galaxies \citep{Volonteri_2003}. It is hard to test this prediction of redshift evolution observationally, since the majority of current observations are at $z < 0.2$ (\autoref{fig:offset_data}). We therefore look forward to the results of applying the methods of \citet{Spingola_2020} to the CLASS sample of lensed sources, since this will probe $z=0.6-3.6$.

\begin{figure}
	\centering
	\includegraphics[width=\columnwidth]{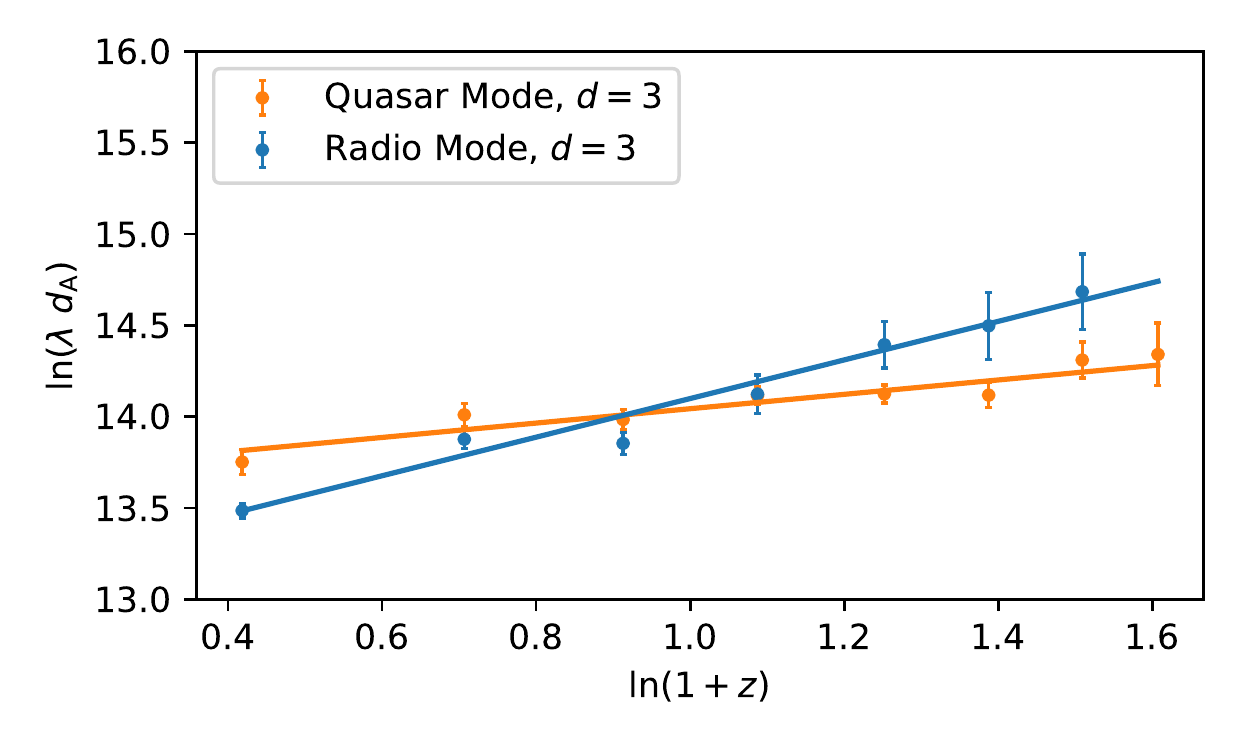}
	 \caption{\label{fig:MC_redshift_evolution_lambda_dA_d_3} Evolution of $\lambda d_{\rm A}$ as a function of redshift, for both the quasar mode and radio mode samples, where $\lambda^{-1}$ characterises the length scale of BH offsets and $d_{\rm A} (z)$ is the angular diameter distance to redshift $z$. Thus $\lambda d_A$ gives the reciprocal of the angular scales over which BHs are intrinsically offset. We plot the best fits to $\lambda d_{\rm A} \propto (1+z)^\alpha$.}
\end{figure}

The most physically interesting parameter from the fits to the offset distributions is $\lambda$, which tells us about the scale to which offset BHs extend. In \autoref{fig:MC_redshift_evolution_lambda_dA_d_3} we plot the redshift evolution of $\lambda d_{\rm A}$, for angular diameter distance $d_{\rm A}$. Fitting this to a power law in the cosmological scale factor, we find 
\begin{equation}
	\lambda d_{\rm A} \propto \left( 1 + z \right)^\alpha, \quad \alpha = 
		\begin{cases}
			\mbox{$0.39 \pm 0.07$,} & \text{Quasar Mode} \\
			\mbox{$1.06 \pm 0.20$,} & \text{Radio Mode}.
		\end{cases}
\end{equation}
A larger value of $\lambda$ indicates that the intrinsically offset BHs reside closer to the galactic centre. Since $\lambda d_{\rm A}$ increases with redshift, we conclude that BHs are more localised to their host's centre at earlier epochs, in terms of observed projected angular offset. This supports the picture of \autoref{fig:h_agn_redshift_evolution} that BHs initially reside near galaxies' centres. 

We note that, although the qualitative trends are the same for $d=2$ and $d=3$, the three-dimensional analysis tends to give a higher probability of a BH being offset than when we use the projected offsets. We find that this is due to a preference of our fitting procedure to obtain larger $\sigma$ for $d=2$ compared to $d=3$. Since our procedure is designed to match the observational technique, this is not a concern provided we only directly compare the observations to the $d=2$ model.

\subsection{Correlation of offsets with galaxy and halo properties}
\label{sec:Correlation of offsets with galaxy and halo properties}

To investigate which parameters besides redshift affect the magnitude of the galaxy-BH offset, we train a Random Forest regressor on the combined quasar plus radio mode sample at $z=0.1$, optimising the regressor's hyperparameters using 5-fold cross-validation. This regressor fits nonlinear decision trees to predict components of the data from others, and is able to provide an estimate of the features most important for the regression \citep{scikit-learn}. Before outlining the results, we describe the features we choose and how we calculate them.

\subsubsection{Chosen features}

From the halo and galaxy finder we obtain the masses of the halo and galaxy, $M_{\rm H}$ and $M_{\rm G}$ respectively, the halo virial radius, $r_{\rm vir}$, and the galaxy's effective radius, $r_{\rm eff}$. We use the black hole mass, $M_{\rm B}$, and Eddington ratio, $\chi$, as directly output from the simulation. From this output we also retrieve the properties of the gas surrounding the BH: the average relative velocity between the BH and gas, $\bar{u}$, the gas density, $\rho_{\rm gas}$, and the average sound speed, $\bar{c}_s$, calculated as described in \autoref{sec:Dynamical friction}. We calculate the mean velocity of the particles in each galaxy, $\bm{v}_{\rm G}$, and hence the relative velocity between the BH and galaxy
\begin{equation}
\bm{v}_{\rm GB} \equiv \bm{v}_{\rm B} - \bm{v}_{\rm G},
\end{equation}
where $\bm{v}_{\rm B}$ is the velocity of the BH. This velocity is decomposed into components parallel, $v_{\rm GB \parallel}$, and perpendicular, $v_{\rm GB \perp}$, to $\bm{r}_{\rm GB}$. We find the angular momentum of the galaxy, $\bm{J}_{\rm G}$, and halo, $\bm{J}_{\rm H}$, in their centre of mass frames, and the corresponding spin parameters \citep{Bullock_2001}, $\lambda_{\rm G}$ and $\lambda_{\rm H}$. The angular momentum of the BH about the galactic centre in the galaxy's centre of mass frame, $\bm{J}_{\rm B}$, is also calculated.

We produce merger trees for all galaxies back to $z=7$, with an average time-step of $\sim 50 {\rm \, Myr}$. Following \citet{Martin_2018}, we identify major mergers to be those with mass ratios, $q_{\rm merge}$, greater than 1:4, where we define the time of the merger, $t_{\rm merge}$, to be the point of coalescence according to the halo-finder, and we calculate the mass ratio at the point at which the less massive galaxy begins to lose mass. The most massive galaxy in the merger has mass $M_1$ at the time $q_{\rm merge}$ is measured. If we do not identify any mergers for a given galaxy fulfilling our mass ratio criterion, then we do not include that system in this part of the analysis. The conclusions of the regression are the same for the other parameters whether or not these systems are included.

To investigate the impact of the external environment we calculate the gravitational field at the BH, $\bm{g}$. We find the density field by applying a cloud-in-cell algorithm to the dark matter and star particles in the simulation, adding this to the gas density field, and solve Poisson's equation on a $512^3$ grid. This corresponds to a minimum spatial resolution of $\sim 200 h^{-1} {\rm \, kpc}$.

Our final features describe the geometry of the system. We include the magnitude of the offset between the halo and galaxy centres, $|\bm{r}_{\rm HG}|$, and the angles between various vectors described above: $\hat{J}_{\rm G}\cdot\hat{J}_{\rm B}$, $\hat{J}_{\rm H}\cdot\hat{J}_{\rm G}$, $r_{\rm GB}\cdot\hat{J}_{\rm G}$, $v_{\rm GB}\cdot\hat{J}_{\rm G}$, $r_{\rm GB}\cdot\hat{g}$, $v_{\rm GB}\cdot\hat{g}$ and $r_{\rm GB} \cdot v_{\rm GB}$, where we denote $\hat{x}$ as the unit vector parallel to $\bm{x}$.

\subsubsection{Correlation results}

\begin{figure*}
	 \includegraphics[width=\textwidth]{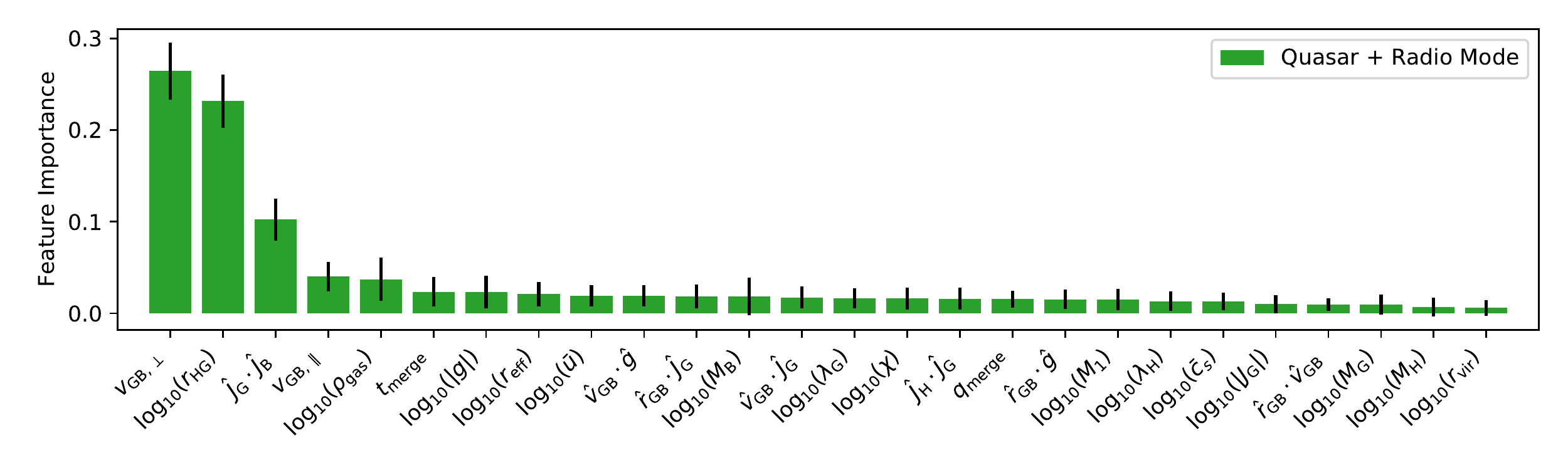}
	  \includegraphics[width=\textwidth]{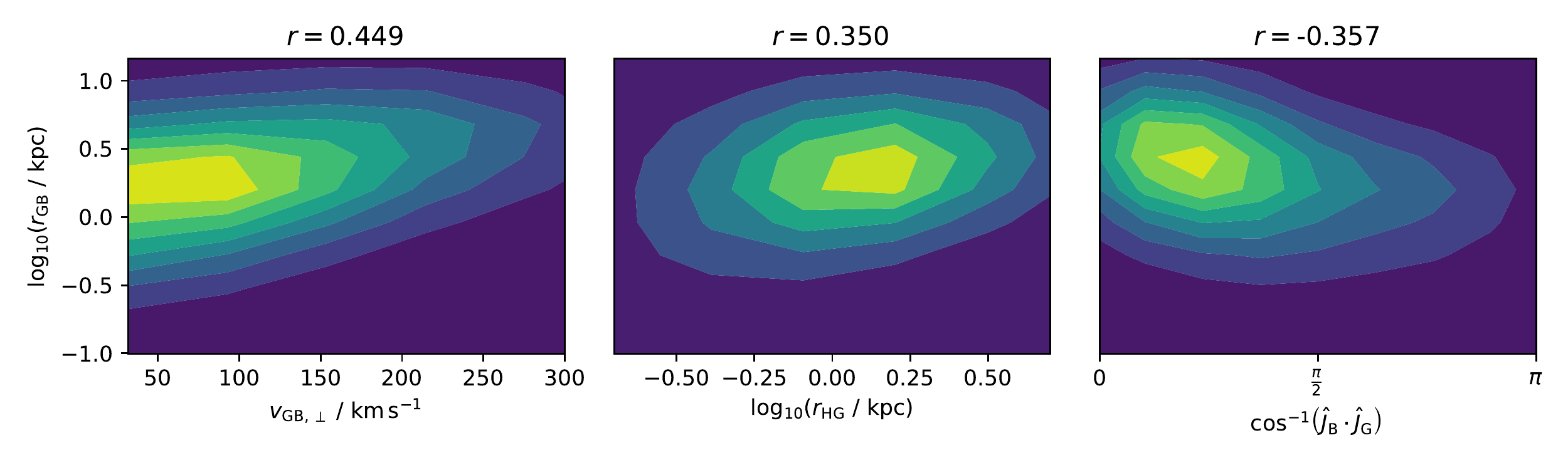}
	 \caption{Feature importances for $\log_{10} ( r_{\rm GB} )$, for BH offset $r_{\rm GB}$, from optimised Random Forests for the combined quasar and radio mode sample from Horizon-AGN at $z=0.1$. The relative velocity between the BH and its host galaxy perpendicular to the offset, $v_{\rm GB, \perp}$, is the most important feature, followed by $r_{\rm HG}$ and $\hat{J}_{\rm G} \cdot \hat{J}_{\rm B}$, with all other parameters relatively unimportant and thus not strongly correlated with $r_\text{GB}$. The two-dimensional histograms for the best three features are shown in the lower panel, with the Spearman's rank correlation coefficients as the titles.}
	 \label{fig:feature_importance_nout_761}
\end{figure*}

The resulting feature importances are plotted in \autoref{fig:feature_importance_nout_761}, where we also plot the two-dimensional distributions of the offsets with the three most importance features: $v_{\rm GB, \perp}$, $r_{\rm HG}$ and $\hat{J}_{\rm G} \cdot \hat{J}_{\rm B}$.

\begin{figure}
	 \includegraphics[width=\columnwidth]{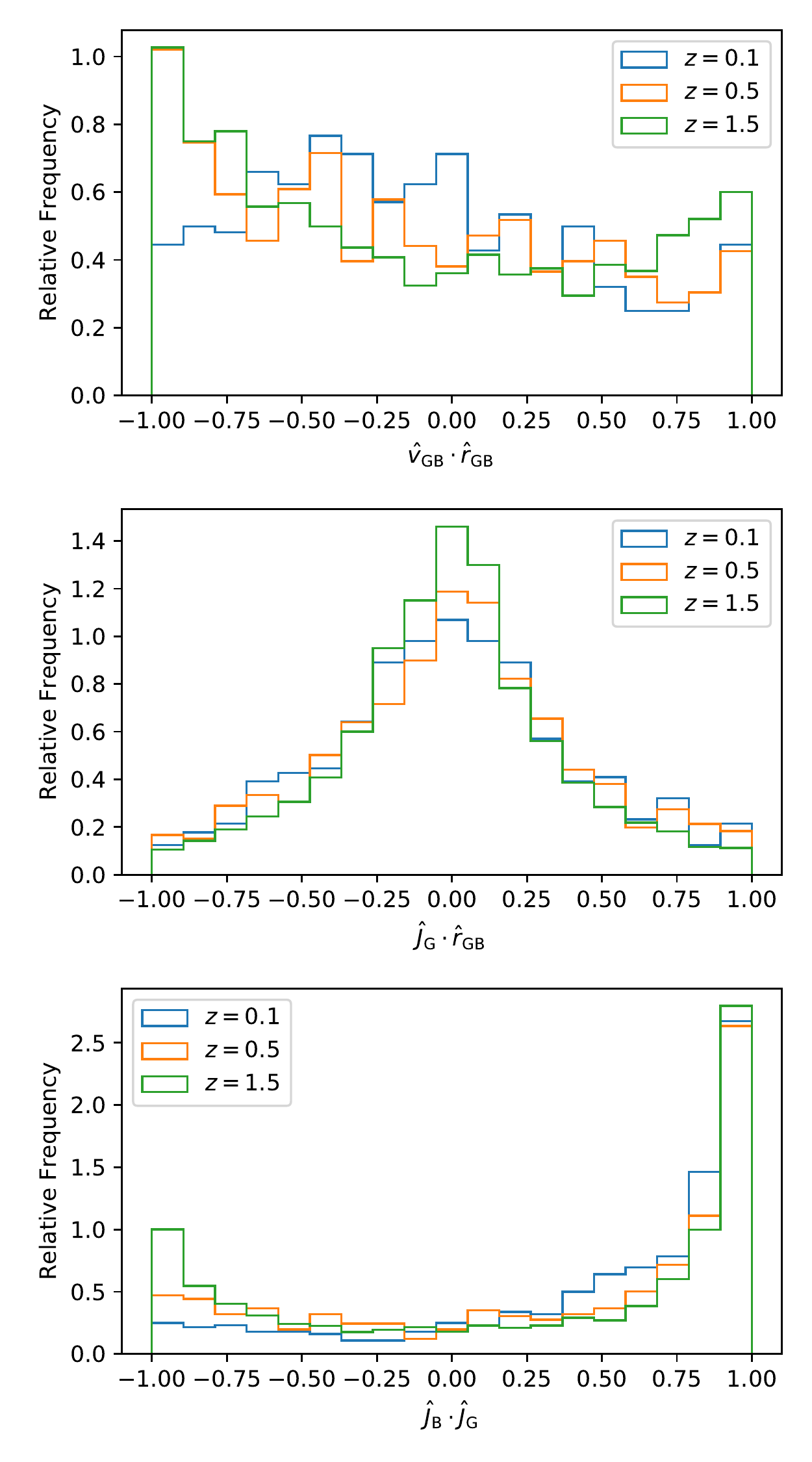}
	 \caption{Distributions of \textit{Upper}: The angle between the BH offset, $\bm{r_{\rm GB}}$, and the velocity of the BH relative to its host galaxy $\bm{v_{\rm GB}}$; \textit{Centre}: The angle between $\bm{r_{\rm GB}}$ and the angular momentum of the galaxy about its centre in its centre of mass frame, $\bm{J}_{\rm G}$; \textit{Lower}: The angle between the angular momentum of the BH about the galactic centre in the galaxy's rest frame, $\bm{J}_{\rm B}$, and $\bm{J}_{\rm G}$. There is a slight propensity for $\bm{r_{\rm GB}}$ and $\bm{v_{\rm GB}}$ to be anti-aligned, so the BH tends to move back towards the galactic centre. This preference is stronger at higher redshift. The BH offsets tend to be perpendicular to $\bm{J}_{\rm G}$, and hence to lie in the plane of the galaxy. $\bm{J}_{\rm B}$ and $\bm{J}_{\rm G}$ are preferentially aligned at all $z$ considered, indicating that BHs move on prograde orbits.}
	 \label{fig:offset_direction_plot}
\end{figure}

We see that the strongest correlation is with $v_{\rm GB, \perp}$, such that BHs with larger offsets have a large $v_{\rm GB, \perp}$. To understand this, in the lower panel of \autoref{fig:offset_direction_plot} we plot the distribution of angles between $\bm{J}_{\rm G}$ and $\bm{J}_{\rm B}$, which we also know is an important feature. Since these tend to align, we conclude that the BHs move on prograde orbits. In particular, 65 per cent of systems are aligned within $60\degr$ at $z=0.1$. Consequently, $v_{\rm GB, \perp}$ gives the orbital velocity of the BH, making it an important parameter for deducing $r_{\rm GB}$.

\begin{figure}
	 \includegraphics[width=\columnwidth]{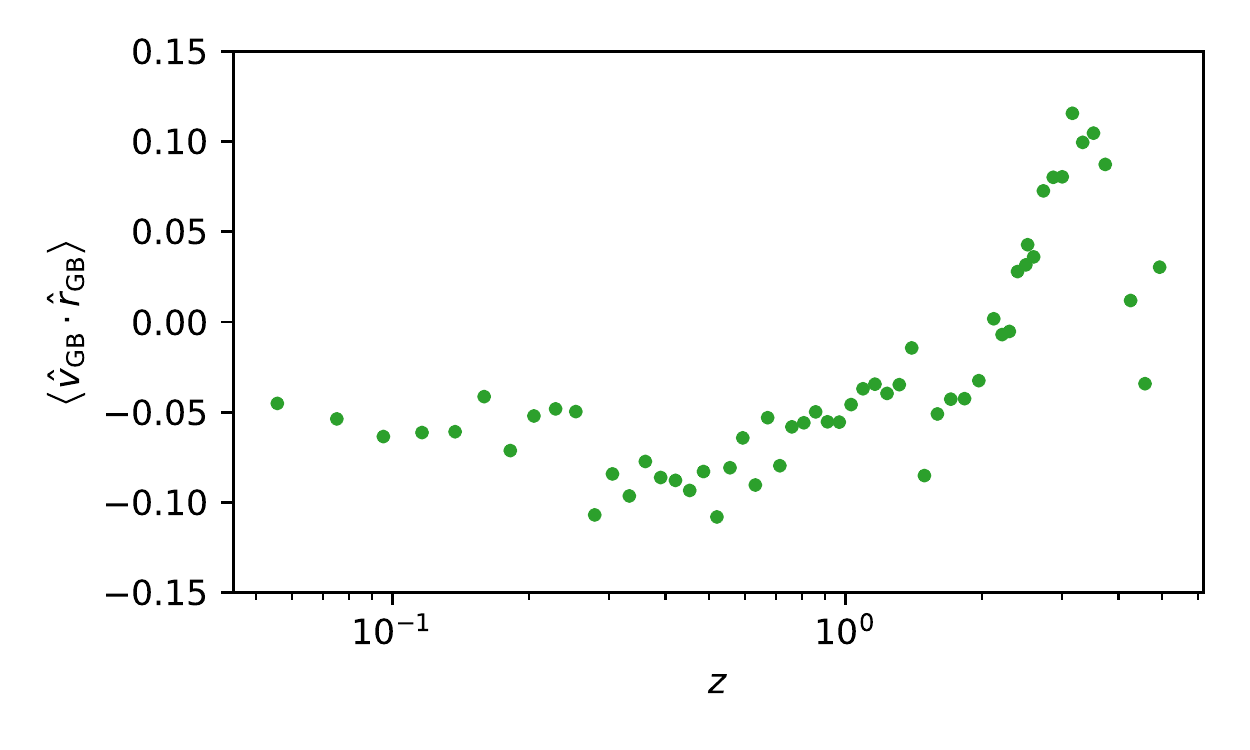}
	 \caption{The mean of $\hat{v}_{\rm GB} \cdot \hat{r}_{\rm GB}$ as a function of redshift, where $\hat{v}_{\rm GB}$ and $\hat{r}_{\rm GB}$ are unit vectors parallel to the relative velocity and position of the BH relative to its host galaxy respectively, for the combined quasar and radio mode sample. For the lowest redshifts, there is a net anti-alignment, so BHs are on average returning back to the galactic centre. The positive value of $<\hat{v}_{\rm GB} \cdot \hat{r}_{\rm GB}>$ at $z=3-4$ corresponds to BHs moving away from the centres of their hosts on average.}
	 \label{fig:mean_vdot_all_z.pdf}
\end{figure}

The two-dimensional distribution of $\cos^{-1}(\hat{J}_{\rm G}\cdot\hat{J}_{\rm B})$ and $r_{\rm GB}$ demonstrates that a wide range of orbital radii are possible if the angular momenta perfectly align, with only smaller offsets permitted as the level of aligned decreases.

In \autoref{fig:offset_direction_plot} we also plot the distributions of the angles between $\bm{r}_{\rm GB}$ and $\bm{v}_{\rm GB}$ for $z=0.1$, 0.5 and 1.5. We plot the mean of these distributions across a wider range of redshift in \autoref{fig:mean_vdot_all_z.pdf}. We see that for $z \lesssim 2$ the velocity of the BH tends to oppose the direction of its offset, i.e. the BHs are, on average, returning to the centres of their host galaxies. The distribution is more uniform at $z = 0.1$ than $z=1.5$. The anti-alignment of $\bm{r}_{\rm GB}$ and $\bm{v}_{\rm GB}$ can be interpreted as the orbital decay due to dynamical friction, which stalls at low $z$, and the uniform part of this distribution is due to the range of eccentricities of the BH orbits. The positive values of $< \hat{v}_{\rm GB} \cdot \hat{r}_{\rm GB} >$ for $z=3-4$ are due to BHs being kicked away from the centres of their galaxies. This is why the fraction of offset BHs is higher today than at the earliest epochs.

We further observe in \autoref{fig:offset_direction_plot} a strong preference for $\bm{r}_{\rm GB}$ to be orthogonal to $\bm{J}_{\rm G}$. If we use $\bm{J}_{\rm G}$ as a proxy for the normal to the plane of the galaxy, we infer that BHs have a propensity to lie in a plane parallel to the galactic plane. We find that 73 per cent of BHs in the combined quasar and radio mode sample have displacements parallel to $\bm{J}_{\rm G}$ within $\pm 1 {\rm \, kpc}$. Given that the resolution of Horizon-AGN is $\sim 1 {\rm \, kpc}$, we therefore conclude that BHs tend to lie in the galactic plane. Horizon-AGN does not contain a procedure for the asymmetric gravitational wave (GW) emission upon BH coalescence and the subsequent recoil, and only contains Schwarzschild BHs. For two Kerr SMBHs in a galaxy merger, if there is a gas rich environment, the spins align with each other and the circumbinary disc's angular momentum \citep{Barausse_2012}. Since gravitational wave emission results in a kick preferentially perpendicular to the BH orbital plane \citep{Lousto_2010}, we would thus expect the distribution to become more isotropic if GW effects were included.

The second most important feature was found to be the magnitude of the halo-galaxy offset, $r_{\rm HG}\equiv |\bm{r}_{\rm HG}|$. This could occur if the halo and BH were tightly bound, and the galaxy is the offset member of the set. However, this is not the case since, as can be seen in the central panel of \autoref{fig:feature_importance_nout_761}, the median halo-galaxy offset is half the median galaxy-BH offset, and the halo-BH and galaxy-BH offsets are well correlated, with a Spearman's rank regression coefficient of $r = 0.81$. Instead, we interpret $r_{\rm HG}$ as a measure of how disturbed the system is; more disturbed systems have larger $r_{\rm HG}$ and thus greater galaxy-BH offsets.

We repeated the analysis using an Extra Trees regressor, separately using the quasar and radio mode samples, using different mass ratios to define a major merger (1:2 and 1:3), choosing the most massive or most recent merger irrespective of $q_{\rm merge}$, and altering when we measure quantities from the merger. In all cases, the results were qualitatively similar, with $v_{\rm GB, \perp}$ always being the most important feature. We also separately analysed the quasar and radio mode samples, and for an uncut Horizon-AGN sample, where we choose the closest BH to the centre of the galaxy, but do not make any cuts on galaxy mass, halo mass, BH luminosity or accretion rate. In all cases, the BH closest to the centre of the galaxy preferentially moves on a prograde orbit in the galactic plane. One difference is that the full Horizon-AGN sample exhibits a slight propensity for $\bm{r}_{\rm GB}$ and $\bm{v}_{\rm GB}$ to be orthogonal at the lowest redshifts, consistent with more circular orbits for the complete sample.

At $z=0.1$, \autoref{fig:feature_importance_nout_761} suggests that the magnitude of the BH offset is independent of $\bm{g}$. We also find that the distribution of the angle between the offset and $\bm{g}$ is consistent with isotropic, with a $p$-value of 0.50 for the two-sided KS test that the distribution is drawn from an isotropic distribution.

\autoref{fig:feature_importance_nout_761} indicates that the quantities describing the merger history of the galaxy are relatively unimportant. Perhaps counter-intuitively, the Spearman correlation coefficient between $\log_{10}(r_{\rm GB})$ and $t_{\rm merge}$ is 0.184, so there is a slight positive correlation. In fact, this is true only for small $t_{\rm merge} \la 5 {\rm \, Gyr}$. Since we define the start of the merger to be the point at which the halo-finder can no longer identify two separate galaxies, there may be some delay between this point and the BHs being dislodged from their centres. Further, if a BH merger event occurs, there is an additional delay due to the finite timescales associated with the binary's evolution \citep{Begelman_1980}, although only a limited period of this evolution will be resolved in Horizon-AGN. Thus, an initial increase of $r_{\rm GB}$ with $t_{\rm merge}$ is not too surprising, and indicates the time required for the central BH to become perturbed. For $t_{\rm merge} \ga 5 {\rm \, Gyr}$ there is little correlation with the offset.

\subsection{Compatibility of observational datasets}
\label{sec:Compatibility of observational datasets}

Before comparing to Horizon-AGN, it is important to investigate the level at which the independent observational datasets agree with one another. To do this we will restrict our attention to samples with more than two systems.

All the datasets that we use include the offsets greater than $3 \sigma$, so an initial test of compatibility is to compare the fraction of systems which have an offset $>3\sigma$. Dividing the systems into two groups (those which have offsets $> 3\sigma$ and those which do not), and assuming a Binomial distribution, with probability $\Lambda$ of a single BH being offset at $>3\sigma$, we can estimate $\Lambda$ and its error using a maximum likelihood estimate.

\begin{figure}
	\centering
	 \includegraphics[width=\columnwidth]{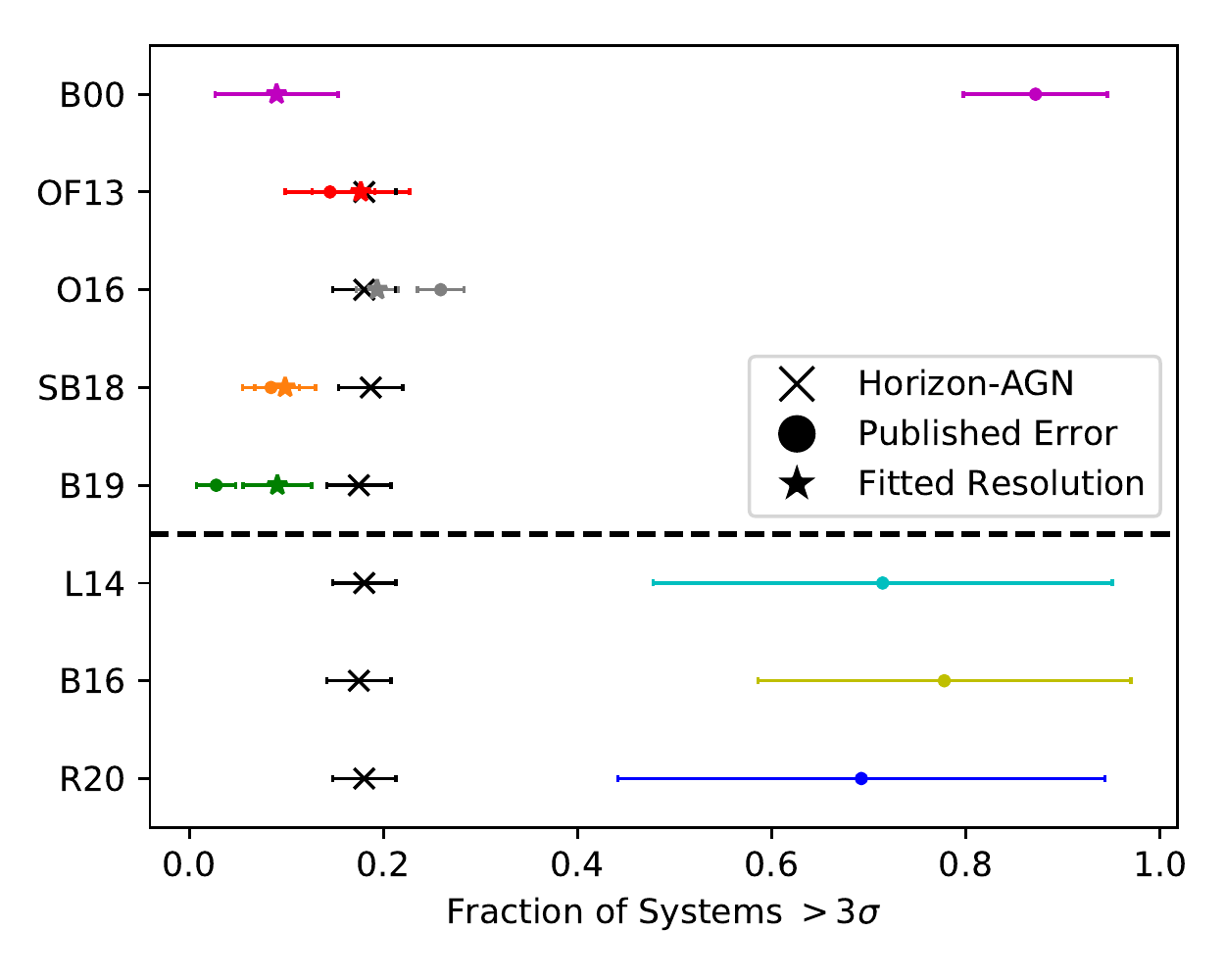}
	 \caption{\label{fig:test_intrinsic_offset_fraction_offset} The fraction of systems with offsets $> 3 \sigma$ from the various datasets (coloured points), compared to different cuts from Horizon-AGN (black points). The errors are from the maximum likelihood estimate and show the 95 per cent confidence intervals. For the circular markers, we set $\sigma$ to be the published error on each offset, whereas for the points marked with stars we use the fitted value of $\sigma$ from \autoref{eq:Offset distribution fit}. The three datasets below the dashed line have very few systems, hence the large errors, and are only included for completeness; we make no further quantitative comparison between these and Horizon-AGN. The reader should therefore focus mainly on the points above the dashed line.}
\end{figure}

We plot the calculated $\Lambda$ and the errors in \autoref{fig:test_intrinsic_offset_fraction_offset}, alongside the values calculated using different cuts from Horizon-AGN. The errors for the Horizon-AGN values are calculated using the errors on the value of $\sigma$, whereas we assume a fixed $\sigma$ for the data. If the sample contains a sufficiently large number of systems, we fit the observational data to \autoref{eq:ad definition} (as described in \autoref{sec:Accounting for finite resolution}) to obtain $\sigma$. Otherwise we used the quoted uncertainty on each offset.

The considerable difference between using the fitted $\sigma$ or the published errors is expected from \citetalias{Binggeli_2000}, since the errors are given as lower bounds, so the offsets relative to these errors will be larger. The inconsistency between the two values of $\Lambda$ for \citetalias{Barrows_2019} suggest that the reported errors may be too conservative, as they give a much lower $\Lambda$ than the fitted offset. To mitigate these issues, henceforth we choose $\sigma$ to be the fitted value for the samples where this is possible (\citetalias{Binggeli_2000}, \citetalias{Orosz_2013}, \citetalias{Orosz_2016}, \citetalias{Skipper_2018} and \citetalias{Barrows_2019}). Otherwise we use the published errors. 

Due to their small sample sizes, the \citetalias{Lena_2014}, \citetalias{Barrows_2016} and \citetalias{Reines_2020} values have very large errors, so it is hard to say whether these are discrepant with the other values. We make no further quantitative comparison using these data, such as assigning a probability of consistency with Horizon-AGN; these are only included for completeness. We see that \citetalias{Binggeli_2000}, \citetalias{Skipper_2018} and \citetalias{Barrows_2019} give similar values of $\Lambda$ of 0.09, 0.10 and 0.09 respectively. These values are lower than those calculated from Horizon-AGN. \citetalias{Orosz_2013} has 18 per cent of systems having offsets greater than $3 \sigma$, which is consistent with the 18 per cent of systems obeying this criterion in Horizon-AGN.

The only data to have a larger fraction of offset BHs than Horizon-AGN is \citetalias{Orosz_2016}, with 19 per cent of systems having offsets above $3\sigma$, although the values are consistent within the errors.

\begin{figure}
	\centering
	 \includegraphics[width=\columnwidth]{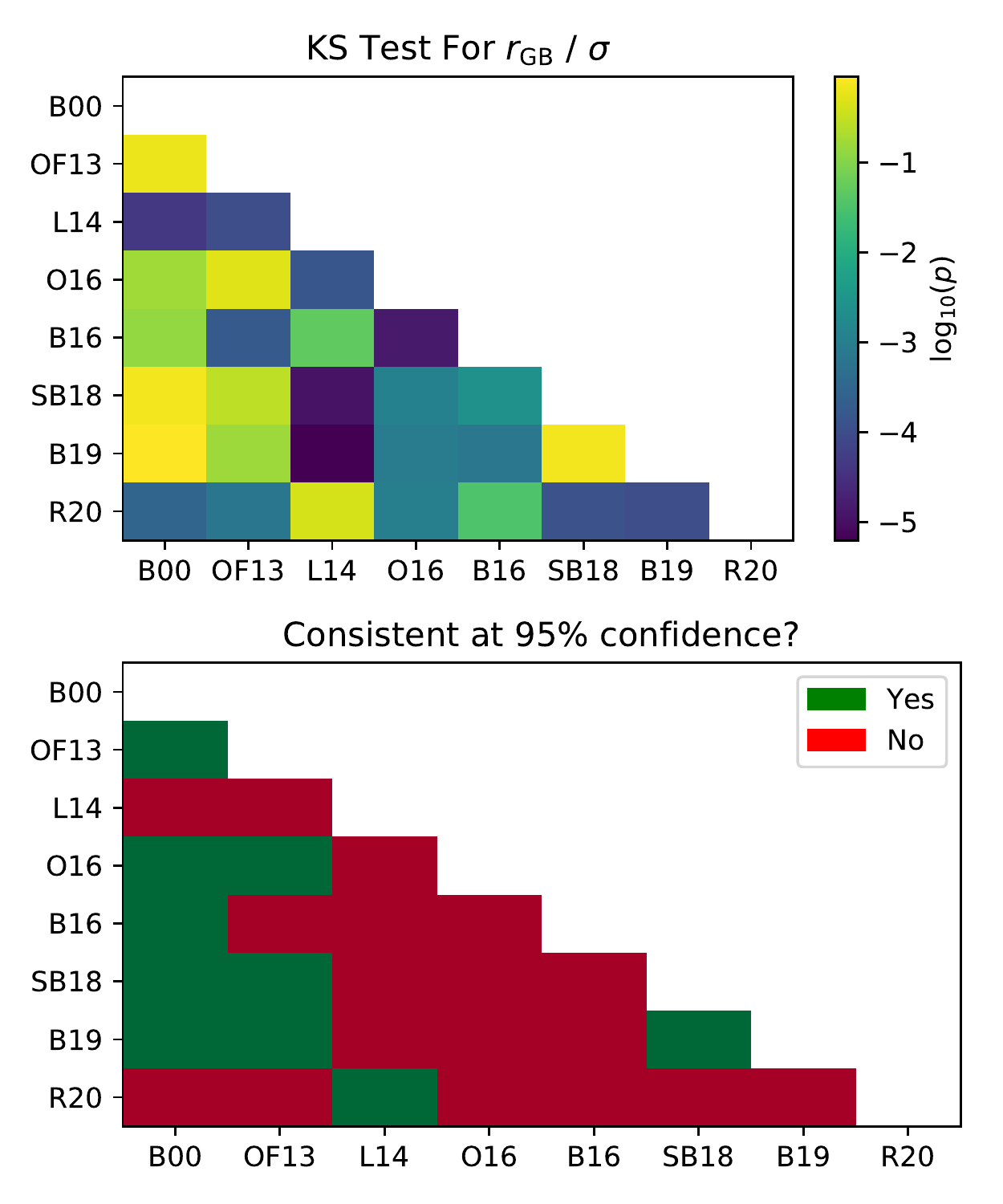}
	 \caption{\label{fig:test_intrinsic_offset_compare_all_datasets} \textit{Upper:} The $p$-values of KS tests between the different datasets' angular offsets, normalised by the resolution, $r_{\rm GB} / \sigma$, to see if they could be drawn from the same distribution. $p=1$ indicates that the underlying distributions are identical. \textit{Lower:} A summary of the KS test, where green indicates $p>0.05$ and red indicates $p<0.05$; the datasets in green are consistent with each other.}
\end{figure}

We now perform a Kolmogorov-Smirnov (KS) test on the normalised angular offset, $r_{\rm GB} / \sigma$, for all the datasets, to see if this normalised offset obeys the same distribution for each sample of galaxies. The results are plotted in \autoref{fig:test_intrinsic_offset_compare_all_datasets}. Since \citetalias{Barrows_2016} only contains the $>3 \sigma$ offsets, we cut the other datasets to include only the $>3 \sigma$ offsets when comparing to \citetalias{Barrows_2016}.

From this we can make the following observations
\begin{itemize}
	\item \citetalias{Binggeli_2000,Orosz_2013,Skipper_2018,Barrows_2019} are mutually consistent with each other.
	\item The $>3 \sigma$ distribution in \citetalias{Binggeli_2000} is consistent with \citetalias{Barrows_2016}.
	\item \citetalias{Lena_2014} is consistent with \citetalias{Reines_2020}.
	\item \citetalias{Orosz_2016} is consistent with \citetalias{Orosz_2013} and \citetalias{Binggeli_2000}.
\end{itemize}
This means that each dataset is consistent with at least one of the others, but the datasets are not all consistent with each other. It is interesting that the tail of \citetalias{Barrows_2019} is not consistent with \citetalias{Barrows_2016}, indicating that the change in selection criteria has altered their results.

\subsection{Comparison of Horizon-AGN with observations}
\label{sec:Comparison of Horizon-AGN with observations}

In \autoref{tab:Pcrit} we give the results for fitting $a_2$ to the various datasets in terms of the fraction of offset BHs in the data, $f_{\rm crit}$, and from Horizon-AGN, $P_2 \left( {\rm offset} \right)$, using \autoref{eq:Offset fraction}. We also calculate the probability of generating datasets with the fraction of offset BHs up to $f_{\rm crit}$, $P_{\rm crit}$, using \autoref{eq:Pcrit} and $N_{\rm dist} \sim 2 \times 10^6$.

As noted in \autoref{sec:Quasar vs radio mode}, we find that the systems appear more offset in three dimensions than if they are projected onto the sky, so for consistency we only consider $d=2$ here. Even using the two dimensional offsets, we find that Horizon-AGN predicts a higher fraction of offset BHs, although the large errors obtained by just using $a_2$ to determine a probability of being offset results in $P_2({\rm offset})$ and $f_{\rm crit}$ being approximately consistent.

In most cases, $P_{\rm crit}$ is never greater than a few per cent, and is only 0.3 per cent for \citetalias{Barrows_2019}, indicating that it is unlikely to create these observations given the offset distributions from Horizon-AGN. The one exception is for \citetalias{Orosz_2016}, where there is a 48 per cent chance of producing up to this fraction of offset BHs. From \cref{fig:test_intrinsic_offset_fraction_offset,fig:test_intrinsic_offset_compare_all_datasets} we know that \citetalias{Orosz_2016} does not agree well with most other observations.
Furthermore, as noted in \autoref{sec:O16}, these data did not undergo follow-up observations, so it is likely that the increased fraction of offset BHs is due to spurious sources. Given that Horizon-AGN overpredicts the fraction of offset BHs compared to the other seven studies considered here, we conclude that the fraction of offset BHs in Horizon-AGN is larger than observed. This result is strengthened by the argument that the calculated value of $f_{\rm crit}$ should be treated as an upper limit (see \autoref{sec:Observational Data}).

\begin{table}
	\caption{\label{tab:Pcrit} Comparison of observations to Horizon-AGN samples ($z=0.1$) designed to mimic their selection criteria. $P_2 ( {\rm offset})$ is the fraction of offset BHs (\autoref{eq:Offset fraction}), calculated by fitting the 2D projected offsets to \autoref{eq:ad definition}. It is independent of the form of the intrinsically offset distribution. $f_{\rm crit}$ is the corresponding quantity for the observations. $P_{\rm crit}$ is the probability of generating a fraction of up to $f_{\rm crit}$ offset BHs. The selection criteria have little effect on $P_2 ( {\rm offset})$; for the quasar-mode sample with no further cuts this is $0.27\pm0.14$.}
	\centering
	\begin{tabular}{*4c}
	{} & $f_{\rm crit}$ & $P_2 \left( {\rm offset} \right)$ & $P_{\rm crit} \left( d = 2 \right)$ \\
	\hline
	OF13 & $0.18 \pm 0.03$ & $0.27 \pm 0.14$ & 0.07 \\
	O16 & $0.24 \pm 0.02$ & $0.27 \pm 0.14$ & 0.48 \\
	SB18 & $0.17 \pm 0.04$ & $0.28 \pm 0.11$ & 0.05 \\
	B19 & $0.09 \pm 0.02$ & $0.27 \pm 0.10$ & 0.003 \\
	\end{tabular}
\end{table}

\section{Discussion}
\label{sec:Discussion}

\subsection{Systematic uncertainties}

\subsubsection{Effect of galaxy mass}
\label{sec:Effect of galaxy mass}

A potentially important consideration is whether the stellar masses of the observed galaxies are compatible with those of the Horizon-AGN sample. The relatively small box size of Horizon-AGN means that it is dominated by lower mass objects. The observational masses are obtained from the MPA-JHU analysis of SDSS data, based on the methods of \citet{Brinchmann_2004}, \citet{Kauffmann_2003} and \citet{Tremonti_2004}.

\begin{figure}
	 \includegraphics[width=\columnwidth]{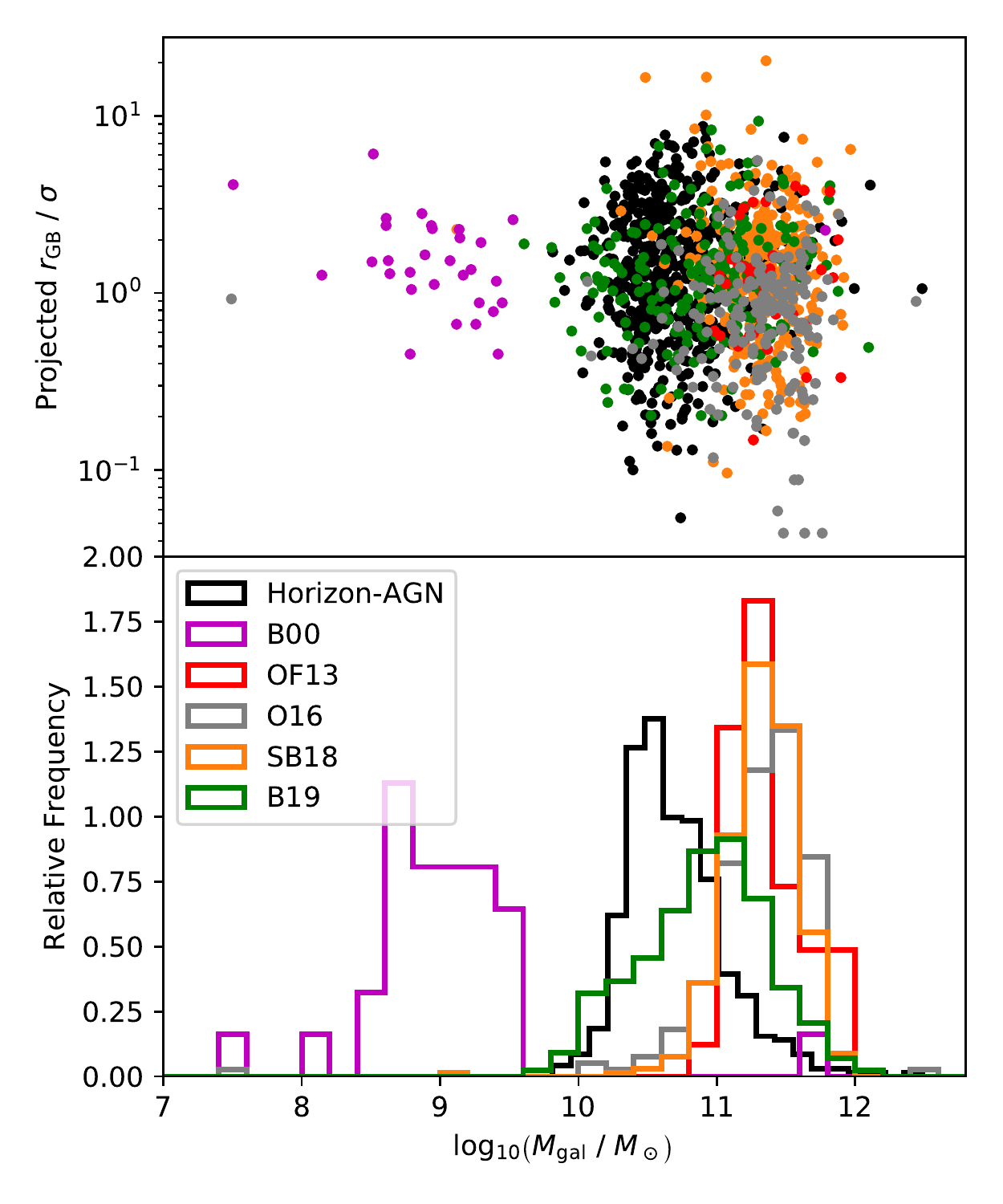}
	 \caption{\label{fig:f_offset_h_agn_mass} The distributions of host galaxy masses considered in \citetalias{Binggeli_2000,Orosz_2013,Orosz_2016,Skipper_2018,Barrows_2019} and the quasar mode sample from Horizon-AGN. In the upper panel we plot the mass-offset plane, where the offsets are normalised by the fitted resolution, $\sigma$.}
\end{figure}

In \autoref{fig:f_offset_h_agn_mass} we plot the mass distributions from the four samples and the mass-offset plane, normalised by $\sigma$. The Spearman regression coefficients and $p$ values for the test that the mass and offset are uncorrelated are given in \autoref{tab:mass correlation}. For all samples except \citetalias{Barrows_2019} there is very little overlap in mass between the observations and Horizon-AGN. However, we find very little variation of offset with galaxy mass, although \citetalias{Barrows_2019} does exhibit a slight positive correlation with $p = 0.03$.

\begin{table}
	\caption{\label{tab:mass correlation} Spearman correlation coefficients, $r$, between the BH offset, $r_{\rm GB}$, and galaxy mass, $M_{\rm G}$. The $p$ value is for a two-sided hypothesis test, with the null hypothesis that $r_{\rm GB}$ and $M_{\rm G}$ are uncorrelated. Most datasets show no significant correlation.}
	\centering
	\begin{tabular}{*3c}
	 Sample & Spearman $r$ & $p$-value  \\
	\hline
	B00 & -0.35 & 0.06 \\
	OF13 & 0.16 & 0.32 \\
	O16 & 0.03 & 0.67 \\
	SB18 & -0.07 & 0.18 \\
	B19 & 0.15 & 0.03 \\
	Horizon-AGN (Quasar Mode) & 0.02 & 0.60 \\
	\end{tabular}
\end{table}

We repeated the Monte Carlo sampling for the \citetalias{Skipper_2018}-like Horizon-AGN sample, but now imposing that the galaxy mass satisfies $M_{\rm G} > 4.3 \times 10^{10} M_\odot$, which gives the most massive 50 per cent of the sample. We find that this mass cut has very little effect on the fitting parameters. In particular, we find that $\mu$ and $s$ are consistent between the two samples; $\mu$ changes from $0.40\pm0.06$ to $0.39\pm0.06$ and $10^3s$ slightly increases from $14\pm2$ to $19\pm3$. This further indicates that, for the mass range considered here, the mass of the galaxy is unimportant. 

\subsubsection{Galactic vs projected luminosity centre}

When considering the offsets from Horizon-AGN, so far we have defined the centre of the galaxy to be the position of the densest star particle within a galaxy, where we use the full three-dimensional information. This clearly is not how the centre is defined in observations, so it is important to check that any discrepancy between observations and simulations is not due to the definition of the centre of the galaxy. To do this we must compare the true position of the galactic centre to that inferred from the projected distribution of star particles.

For every galaxy we find all star particles within a box, $\mathcal{B}_{\rm G}$, of side length $4 r_{\rm eff}$, centred on the galactic centre. The projected coordinates of the $i^{\rm th}$ star particle are
\begin{equation}
	x_i = \left( \alpha_i - \alpha_{\rm G} \right) \cos \delta_{\rm G}, \qquad y =  \delta_i - \delta_{\rm G}
\end{equation}
where the projected true centre has right ascension $\alpha_{\rm G}$ and declination $\delta_{\rm G}$ and the star particle has right ascension $\alpha_i$ and declination $\delta_i$, measured by an observer at the centre of the box. Assuming these are distributed according to a S\'{e}rsic \citep{Sersic_1963} distribution, such that the probability of being at some coordinates $(\tilde{x}_i, \tilde{y}_i) = (x_i - x_0, y_i - y_0)$, is
\begin{equation}
	p \left( \tilde{x}_i, \tilde{y}_i \right) = I_0 \exp \left( - b_n \left[ \left( \frac{R_i}{R_{\rm eff}} \right)^{\frac{1}{n}} - 1 \right] \right),
\end{equation}
where $\tilde{x}_i = R_i \cos \varphi$ and $\tilde{y}_i = R_i \left( 1 - \epsilon \right) \sin \varphi$, for polar angle $\varphi \in [0, 2\pi )$ relative to the major axis, which is at an angle $\theta$ relative to the $x$ axis, and ellipticity $\epsilon \in [0, 1)$. We normalise the probability distribution,
\begin{equation}
	I_0 = \frac{b_n^{2n} e^{-b_n}}{2 \pi n R_{\rm eff}^2 \Gamma \left( 2n \right) \left( 1 - \epsilon \right)},
\end{equation}
and define $R_{\rm eff}$ such that half of the probability lies within $R_{\rm eff}$, 
\begin{equation}
	\frac{\gamma \left( b_n, 2n \right)}{\Gamma \left( 2n \right)} = \frac{1}{2},
\end{equation}
where $\gamma \left(z, a \right)$ is the incomplete lower gamma function. We thus must fit for 6 parameters ($\{ R_{\rm eff}, n,  x_0,  y_0,  \epsilon, \theta \}$), which we do by maximising the likelihood
\begin{equation}
	\log \mathcal{L}_{\rm G} \left( R_{\rm eff}, n,  x_0,  y_0,  \epsilon, \theta \right) = \sum_{i \in \mathcal{B}_{\rm G}} \log p \left( \tilde{x}_i, \tilde{y}_i \right).
\end{equation}

We impose uniform priors on all parameters, so that the maximum likelihood is also the maximum of the posterior. We run the optimisation 5 times for each galaxy, with a different start point each time, generated randomly from the priors and adopt the maximum likelihood of these 5 as the true maximum likelihood. We find this number to be sufficient for several of the endpoints of the optimiser to be coincident at a likelihood value above the remainder, indicating that they have reached the maximum-likelihood point and therefore that 5 repeats is sufficient.

Using the fitted projected luminosity centres $(x_0, y_0)$, and assuming that the radial distance to the galactic centre is the same in this method as before, we can thus find the coordinates of the observed luminosity centre, $\bm{r}_{\rm G, lum}$.

Working with the quasar mode sample at $z=0.1$, in 71 per cent of cases, the luminosity centre shifts by less than $1 {\rm kpc}$. With both $d=2$ and $d=3$, the mean BH offset shifts by less than $0.08 {\rm kpc}$. Fitting the results to $a_d$, we find that probabilities of being offset is
\begin{equation}
	\begin{split}
		P_2 \left( {\rm offset, \ lum}\right) &= 0.30 \pm 0.12 \\
		P_3 \left( {\rm offset, \ lum}\right) &= 0.40 \pm 0.16,
	\end{split}
\end{equation}
which are consistent with the values from the galactic centre in \autoref{tab:Pcrit}.

The effects are slightly greater for the radio mode sample, where 56 per cent of cases shift by less than $1 {\rm \, kpc}$ and the mean difference is $0.18 {\rm \, kpc}$. This is still smaller than the resolution of Horizon-AGN. The probabilities of being offset in two and three dimensions for the radio mode BHs are
\begin{equation}
	\begin{split}
		P_2 \left( {\rm offset, \ lum}\right) &= 0.31 \pm 0.09 \\
		P_3 \left( {\rm offset, \ lum}\right) &= 0.402 \pm 0.10,
	\end{split}
\end{equation}
which are again consistent with \autoref{tab:Pcrit}.

We therefore conclude that the offset fraction in Horizon-AGN does not change significantly if we use the luminosity, rather than galactic, centre.

\subsection{Comparison with other simulations}

The majority of cosmological hydrodynamical simulations besides Horizon-AGN do not include a prescription for dynamical friction on BHs. For example, Illustris-TNG \citep{Weinberger_2017}, MassiveBlack-II (\citealt{Khandai_2015}, Khandai private communication) and Simba \citep{Dave_2019} artificially return their BHs back to the potential minimum of their host. The EAGLE simulations \citep{Schaye_2015} also do this for BHs with masses below 100 times the gas particle mass.

The Magneticum simulations, however, do attempt to keep their BHs near the galactic centres using physical processes \citep{Hirschmann_2014}: they impose strict momentum conservation during gas accretion and BH mergers and they include Chandrasekhar friction \citep{Chandrasekhar_1943}. The peak of their distribution of the BH offsets relative to the host's potential minimum occurs at $0.7 h^{-1} {\rm \, kpc}$. Assuming this distribution takes the form of $a_d$ (\autoref{eq:ad definition}) with $d=3$, we would envisage the peak to be at $r_{\rm GB} = \sqrt{(d-1)} \sigma$. Taking $\sigma$ to be half the minimum gravitational softening length ($2.0 h^{-1} {\rm \, kpc}$), as is approximately true for Horizon-AGN, this is value we would predict. Thus the distributions in offsets between the simulations are qualitatively similar and dominated by the resolution.

In \autoref{sec:Correlation of offsets with galaxy and halo properties}, we found that offset BHs tend to lie within the galactic plane. This is in contrast with \citet{Tremmel_2018}, who found that wandering BHs are preferentially found outside of galactic discs, with $>4\sigma$ confidence, for Milky Way (MW)-type haloes. To make a more meaningful comparison with \citet{Tremmel_2018}, we define a MW-type halo as having a mass $5 \times 10^{11} \leq M_{\rm H} / {\rm M_{\sun}} \leq 2 \times 10^{12}$. We consider all MW-type haloes which are in a galaxy+halo structure and find all wandering BHs within $10{\rm \, kpc}$ of the centre of the halo. We use \citet{Tremmel_2018}'s definition of a wandering BH as one further than $0.7{\rm \, kpc}$ from the halo centre. Defining the galactic plane such that the normal of the plane is aligned with $\bm{J}_{\rm G}$, we find that $50$ per cent of BHs lie within $30\degr$ of the galactic plane (which is the value expected for a uniform distribution), compared to $20\pm7$ per cent for \citeauthor{Tremmel_2018}'s sample. Comparing this distribution to an isotropic distribution with a KS-test, we obtain a $p$-value of $p=0.50$, indicating that wandering BHs in MW-type haloes in Horizon-AGN are distributed isotropically.

These results are not necessarily in tension with those of \citeauthor{Tremmel_2018}, since we analyse a very different population of BHs: the minimum mass BH of \citeauthor{Tremmel_2018} is $10^6 {\rm \, M_{\sun}}$, whereas we cannot use BHs below $2 \times 10^7 {\rm \, M_{\sun}}$. Since the wandering BH population only dominates the mass budget of BHs below $\sim 10^5 {\rm \, M_{\sun}}$ \citep{Volonteri_2003}, we do not expect to capture the behaviour of the full wandering BH population in Horizon-AGN, but only the most massive cases. This could also explain why we find far fewer BHs within the virial radius of a MW-type halo than \citeauthor{Tremmel_2018}: $1.4 \pm 0.7$ compared to $12.2 \pm 8.4$.

As noted in \autoref{sec:Accounting for finite resolution}, the large value of $\sigma$ for Horizon-AGN means that we have to introduce a sigmoid function to suppress the contribution of $b_d$ at small offsets. The upcoming New Horizon simulation (Dubois et al., in preparation) is a zoom-in of a `field' environment of Horizon-AGN, with a comoving radius of $10 {\rm \, Mpc}$. The simulation has reached $z=0.7$ at the time of writing and has a much higher resolution than Horizon-AGN, with a maximum spatial resolution corresponding to a physical scale of $40 {\rm \, pc}$ at $z=0$. It will be interesting to see how our results change with this increased resolution, potentially removing the need for the sigmoid suppression and thus making the marginal distribution of $b_d$ equal to $b_{d-1}$.

\subsection{Interpretation of results}

The dynamical friction model used in Horizon-AGN results in a resolution-dominated distribution of offsets of BHs from the centres of their host galaxies. At large $r_{\rm GB}$ we find an exponentially decaying tail, as observed. The fraction of offset BHs in Horizon-AGN, however, is larger than is observed, with a probability of only a few per cent of generating the observations from the simulated distributions (\autoref{tab:Pcrit}).

The dynamical friction model used (\autoref{eq:Dynamical friction}) only depends on the gas parameters, and not on the star or dark matter particles. In order to resolve the force from these particles, the resolution should obey \citep{Pfister_2017}
\begin{equation}
	\Delta x < \frac{GM_{\rm B}}{\sigma_v^2} = 1 {\rm \, pc} \left( \frac{M_{\rm B}}{10^7 {\rm \, M_{\sun}}} \right) \left( \frac{\sigma_v}{200 {\rm \, km \, s^{-1}}} \right)^{-2},
\end{equation}
where $\sigma_v$ is the velocity dispersion of the particles of interest. Remembering that $\Delta x \sim 1 {\rm \, kpc}$ for Horizon-AGN, we see that it is not inconceivable that we would fail to obey this criterion, so that to improve the offset prediction a sub-grid model for dynamical friction from these particles would be needed.

This was investigated by \citet{Pfister_2019}, who found that the stellar component of dynamical friction is more stabilising than its gaseous counterpart, ensuring that BHs remain centralised post-merger. At higher redshifts, however, the irregular stellar distribution prevents stars from providing this constant acceleration.

Although the current dynamical friction model has the desired effect of providing a restoring force such that the BH velocity tends to oppose its displacement (\autoref{fig:offset_direction_plot}), this orbital decay appears to stall at low redshift. Given that this is where \citeauthor{Pfister_2019} find stellar dynamical friction to be most effective, our results indicate that such a model should be included to match the observed BH offset distribution more closely.

It would be more concerning if the fraction of offset BHs in Horizon-AGN was smaller than observed, since reasonable alterations must increase the dynamical friction force and thus reduce the magnitudes of the offsets. The contributions from stars and dark matter could not push a BH to larger $r_{\rm GB}$.

\section{Conclusions}
\label{sec:Conclusions}

We have studied the statistical properties of the offsets between BHs and the centres of their host galaxies at a range of redshifts in the Horizon-AGN simulation, and compared to a set of observations. The Horizon-AGN simulation is almost unique among cosmological hydrodynamical simulations in employing a subgrid model of dynamical friction for BHs rather than artificially advecting them to galaxies' centres at each timestep.

We described the distribution of these offsets as the combination of a Gaussian and a generalised symmetric Laplace distribution, where a more intrinsically offset BH has a larger contribution from the latter. The fraction of offset BHs was compared to observations from the literature, where we tested for consistency and investigated the relative mass dependences of the offsets. From a feature importance analysis we determined the properties of the halo-galaxy-BH system that most strongly affect the offset in the simulation and hence derived a physical picture for the system's evolution.

Our key findings are as follows:

\begin{itemize}
    \item \textit{The fraction of intrinsically offset BHs is higher in the simulation than in most observations}. Although both distributions are dominated by the resolution, we find $\sim 27$ per cent of BHs in Horizon-AGN are intrinsically offset compared to the upper limits of $\sim 10 - 20$ per cent in the observations. We believe this to be due to the unmodelled dynamical friction from stars and dark matter in the simulation.
    \item \textit{A higher fraction of the simulated BHs are intrinsically offset today than at earlier epochs}. 
    This suggests that BHs form near the centres of galaxies at early times, but are then displaced by interaction and mergers, before slowly migrating back. This is expected in the hierarchical structure formation paradigm, where galaxies and their BHs interact and merge to form larger ones.
    \item \textit{Offset BHs in Horizon-AGN exist on prograde orbits in the plane of the galaxy with orbital radii that decay over time}. Although the total fraction of offset BHs increases with time, a given BH moves back towards the centre of its host galaxy due to dynamical friction, unless something causes it to become more offset. This orbital decay stalls at low redshift.
\end{itemize}

To prevent the orbital decay of BHs from stalling at low redshift, and hence align the fraction of offset BHs from the simulation more closely with the observations, future simulations should model the dynamical friction from stars and dark matter as well as gas. With observations planned to probe higher redshifts \citep{Spingola_2020}, it will soon be possible to compare the offset distributions at an epoch when dynamical friction from stars is truly subdominant to that from gas, providing a further test of the BH physics that is implemented in cosmological simulations. Ultimately, this will be necessary to understand fully the role played by SMBHs in shaping the galaxy population over cosmic time.

\section*{Acknowledgements}

We would like to thank the Horizon-AGN collaboration for allowing us to use the simulation data, and particularly Stephane Rouberol for smoothly running the Horizon Cluster hosted by the Institut d'Astrophysique de Paris where most of the processing of the raw simulation data was performed. Some of the numerical work also made use of the DiRAC Data Intensive service at Leicester, operated by the University of Leicester IT Services, which forms part of the STFC DiRAC HPC Facility (\url{www.dirac.ac.uk}). The equipment was funded by BEIS capital funding via STFC capital grants ST/K000373/1 and ST/R002363/1 and STFC DiRAC Operations grant ST/R001014/1. DiRAC is part of the National e-Infrastructure. We are very grateful to Scott Barrows, Gabor Orosz and Sandor Frey for providing data necessary for this work. We also thank Garreth Martin for assistance in constructing merger trees for Horizon-AGN.

DJB is supported by STFC. HD is supported by St John's College, Oxford, and acknowledges financial support from ERC Grant No 693024 and the Beecroft Trust. PGF is supported by the ERC, STFC, and the Beecroft Trust. The research of AS and JD is supported by the Beecroft Trust and STFC.  

\section*{Data availability}

The data underlying this article were provided by the Horizon-AGN collaboration, Scott Barrows, Gabor Orosz and Sandor Frey by permission. Data will be shared on request to the corresponding author with permission of the Horizon-AGN collaboration, Scott Barrows, Gabor Orosz and Sandor Frey. All other data can be retrieved from the references in \autoref{tab:data_summary}.

\bibliographystyle{mnras}
\bibliography{references}

\bsp	
\label{lastpage}
\end{document}